\documentclass[preprint,journal]{vgtc}   
\pdfoutput=1    
\usepackage[final]{changes}

\graphicspath{{figs/}{figures/}{pictures/}{images/}{Figures/}{./}} % where to search for the images

\usepackage{times}  % I think this is now in the style sheet
\usepackage{textcomp}
\usepackage{mathptmx}                  % use matching math font

\usepackage{cite}      % needed to automatically sort the references

\usepackage{xspace}
\usepackage{circledsteps}
\PassOptionsToPackage{svgnames}{xcolor}
\usepackage[svgnames]{xcolor} %  this is part of the IEEE setup - but put here to be clearer
\usepackage{adjustbox}
\usepackage{pdfpages}

\definecolor{NavyBlue}{rgb}{0, 0, 0.5}

\newcommand{\sysname}{AbstractsViewer\xspace}

\newcommand{\circlet}[1]{\Circled[fill color=red, outer color=red, inner color=white]{\textsf{\textbf{#1}}}}
\newcommand{\example}[1]{\circlet{example ref}}

\newcommand{\mgcomment}[1]{}
\newcommand{\etal}{\textit{et al.}}

\newcommand{\cut}[1]{}

\newcommand{\term}[1]{\texttt{#1}\xspace}

\newcommand{\particip}[1]{\textsf{P#1}}
\newcommand{\pquote}[1]{``#1''}

\newcommand{\posthoc}{post hoc\xspace}

\newcommand{\multiscale}{multiscale\xspace}

\newcommand{\designchoice}{\textbf{Design Choice}\xspace}

\newcommand{\SearchTools}{\textsf{Search Tools Panel}\xspace}
\newcommand{\SearchList}{\textsf{Search List}\xspace}
\newcommand{\MapView}{\textsf{Corpus Map}\xspace}
\newcommand{\CorpusMap}{\textsf{Corpus Map}\xspace}
\newcommand{\RegionScatterplot}{\textsf{Region Scatter Plot View}\xspace}
\newcommand{\RegionMat}{\textsf{Region Matrix View}\xspace}
\newcommand{\RegionList}{\textsf{Region List}\xspace}
\newcommand{\NeighborMat}{\textsf{Neighborhood Matrix View}\xspace}
\newcommand{\DocView}{\textsf{Document View}\xspace}
\newcommand{\NeighborList}{\textsf{Neighbor List View}\xspace}
\newcommand{\RadialPlot}{\textsf{Radial Neighborhood View}\xspace}

\newcommand{\Tikzcirclet}[3]{
    \draw (#2,#3) node[circle, fill=red, inner sep=0pt, minimum size=.35cm]{};
    \node[text=white] (A) at (#2,#3) {\fontfamily{phv}\selectfont\textbf{\small #1}};
}

\newcommand{\LargeTikzcirclet}[3]{
    \draw (#2,#3) node[circle, fill=red, inner sep=0pt, minimum size=.45cm]{};
    \node[text=white] (A) at (#2,#3) {\fontfamily{phv}\selectfont\textbf{\large #1}};
}

\newcommand{\DynamicTikzcirclet}[3]{
  \draw (#2, #3) node[anchor=west] {\fontfamily{phv}\selectfont\textbf{\small \Circled[fill color=red, outer color=red, inner color=white]{\textsf{\textbf{#1}}}}};
}

\newcommand\teaserfig{
    \centering

    \begin{tikzpicture}
        \node[anchor=south west,inner sep=0] (image) at (0,0) {\includegraphics[width=\textwidth, trim=0 20mm 0 0,clip]{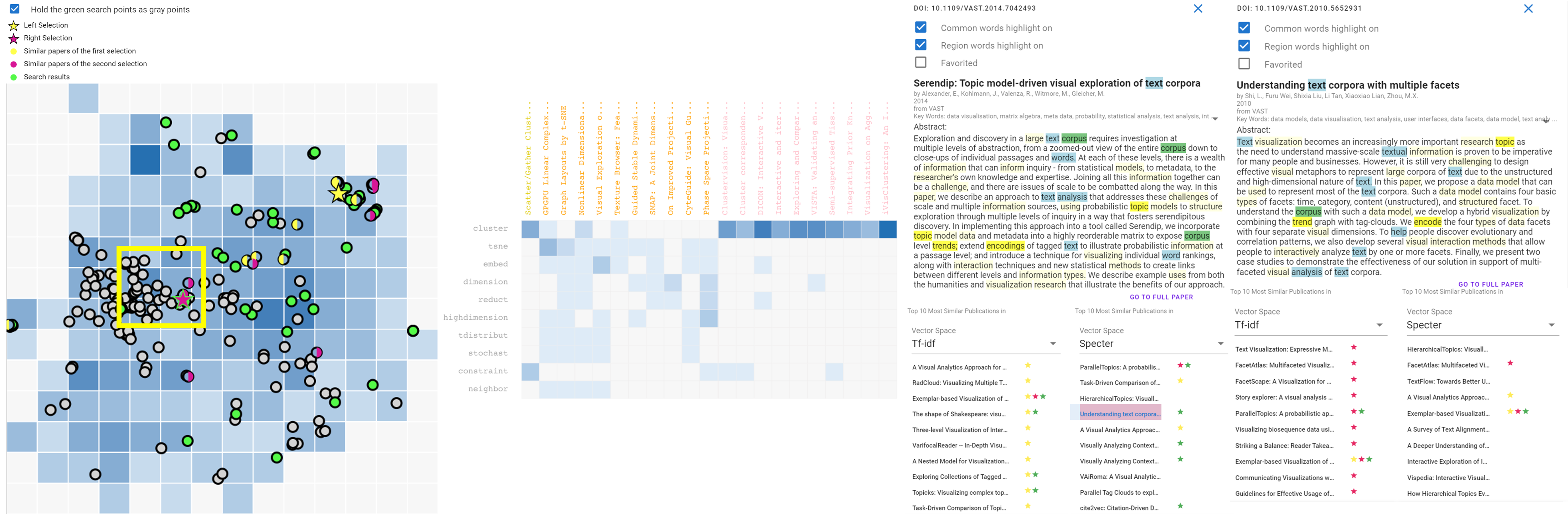}};
        \begin{scope}[x={(image.south east)},y={(image.north west)}]
           
            \LargeTikzcirclet{A}{.25}{.9}
            \LargeTikzcirclet{B}{.55}{.9}
            \LargeTikzcirclet{C}{.95}{.9}
            \end{scope}
    \end{tikzpicture}
    \vspace{-.25in}
    \caption{
        Our approach enhances standard text corpus exploration views with post hoc explanations and support for comparison. 
        \circlet{A} An embedding-based corpus map is shown as a gridded heatmap with circle overlays for search results. This view is enhanced with explanations of region contents (either by hovering over a heatmap square or selecting an arbitrary region shown in yellow),
        the ability to compare two searches (green and gray circles), and two selected documents (pink and yellow stars) allowing their neighbors to be compared (pink and yellow circles).
        \circlet{B} A term-document matrix view is enhanced with salience functions that reorder it to emphasize subsets that explain selected groups. Comparative features  highlight differences between sets of documents.
        \circlet{C} A text view is enhanced with comparison features to show two selected documents. Each document view can highlight explanations for why the document is in its map regions (blue) and why the documents may be considered similar (yellow). Each document provides its most similar neighbors in two vector spaces, with colored symbols to enable comparison between lists. 
      \label{fig:teaser}
    }
}
\newcommand\teaserasfig{
    \begin{figure*}
        \teaserfig
    \end{figure*}
}

\newcommand\system{
\begin{figure}
  \centering
  \begin{tikzpicture}
            \node[anchor=south west,inner sep=0] (image) at (0,0) {\includegraphics[width=0.5\textwidth]{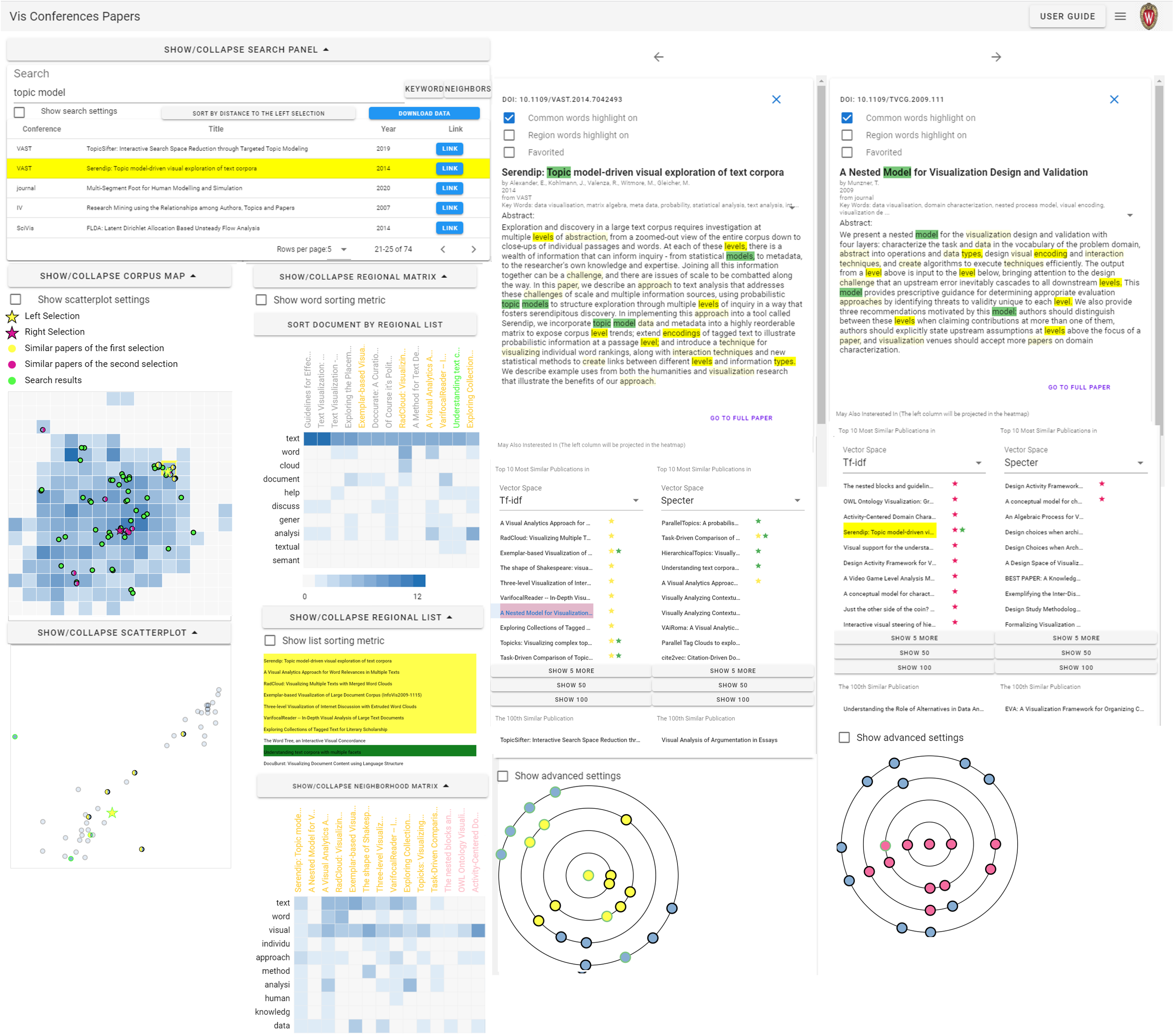}};
            \begin{scope}[x={(image.south east)},y={(image.north west)}]
               
                \Tikzcirclet{A}{.025}{.955}
                \Tikzcirclet{B}{.025}{.6}
                \Tikzcirclet{C}{.025}{.35}
                \Tikzcirclet{D}{.23}{.65}
                \Tikzcirclet{E}{.23}{.37}
                \Tikzcirclet{F}{.23}{.15}
                \Tikzcirclet{G1}{.44}{.94}
                \Tikzcirclet{G2}{.72}{.94}
                \Tikzcirclet{H1}{.48}{.53}
                \Tikzcirclet{H2}{.76}{.56}
                \Tikzcirclet{I1}{.44}{.23}
                \Tikzcirclet{I2}{.72}{.26}
          
                \end{scope}
            
    \end{tikzpicture}
  \caption{\label{fig:system}
    Screenshot of \emph{\sysname} showing its views described in \autoref{sec:views}:
    \circlet{A} \SearchTools including the \SearchList,
    \circlet{B} \MapView,
    \circlet{C} \RegionScatterplot, 
    \circlet{D} \RegionMat,  
    \circlet{E} \RegionList,
    \circlet{F} \NeighborMat,
    \circlet{G} \DocView,
    \circlet{H} \NeighborList,
    and
    \circlet{I} \RadialPlot.
    Two of G, H and I are shown, one for each selection.
  }
\end{figure}
}

\newcommand\awarenessimage{
    \begin{figure*}
    \centering
      \begin{tikzpicture}
            \node[anchor=south west,inner sep=0] (image) at (0,0) {\includegraphics[width=\textwidth]{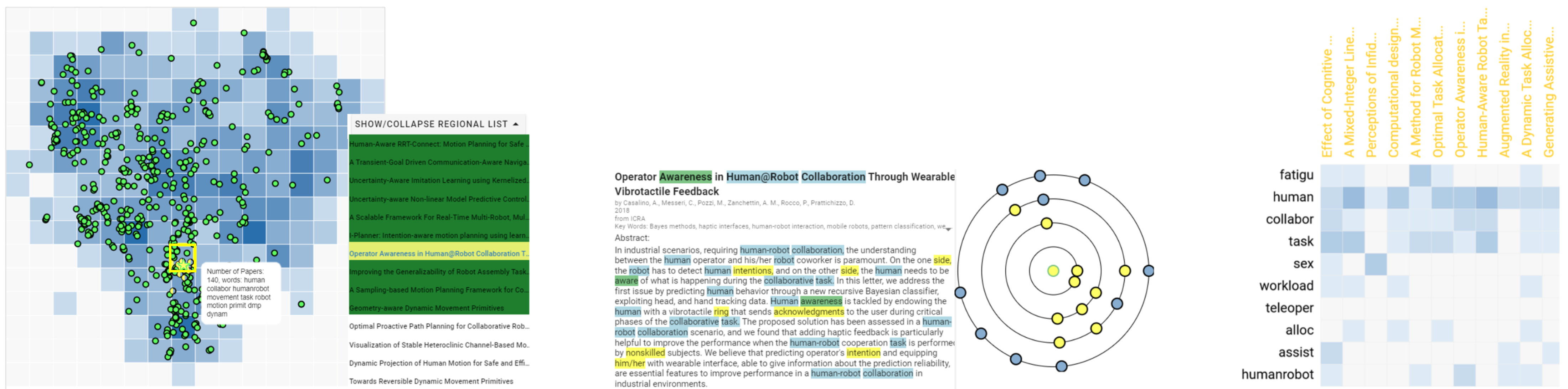}};
            \begin{scope}[x={(image.south east)},y={(image.north west)}]
              
              \LargeTikzcirclet{A}{.025}{.025}
              \LargeTikzcirclet{B}{.4}{.025}
              \LargeTikzcirclet{C}{.8}{.025}
        
              \end{scope}
        \end{tikzpicture}
 
        \caption{\label{fig:awareness}
          Exploration for ``human awareness of robots'' described in \autoref{sec:awareness}. \circlet{A} A keyword search for \term{awareness} yields many hits (green dots in the \CorpusMap). Hovering over regions reveals a region promising region (yellow square). Sorting the \RegionList reveals papers in the region that seem promising. We select a paper. \circlet{B} \DocView shows that the abstract connects to the region (blue highlights). The \RadialPlot shows its neighbors form groups. \circlet{C} The neighborhood of the second document discovered shows that its neighbors connect to it through a diverse set of topics, other than awareness.
        }
      \end{figure*}
    }

\newcommand\viewsfig{
    \begin{figure*}
    \centering
      \begin{tikzpicture}
            \node[anchor=south west,inner sep=0] (image) at (0,0) {\includegraphics[width=\textwidth]{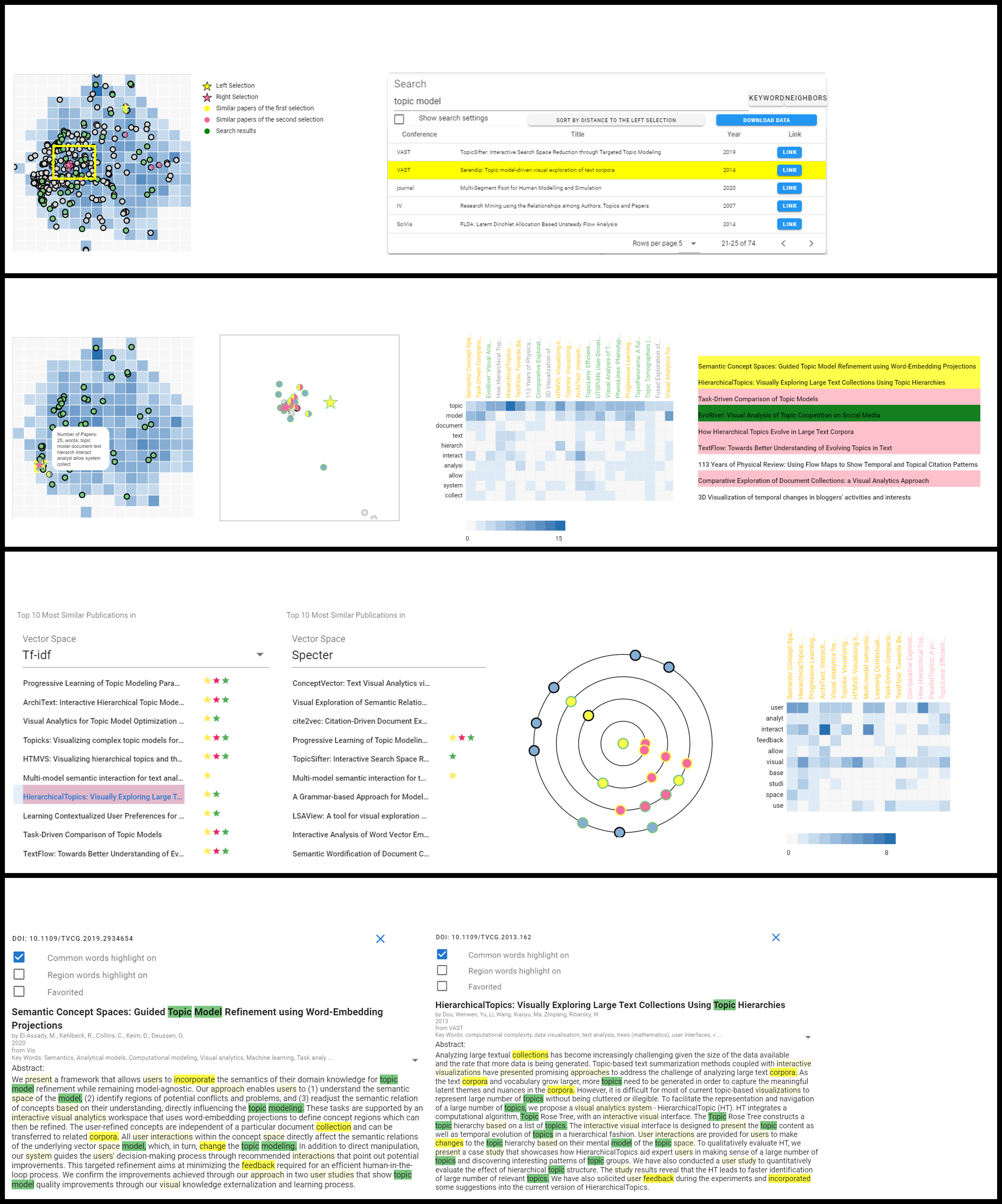}};
            \begin{scope}[x={(image.south east)},y={(image.north west)}]
              
                \draw (.155,.97) node {\LARGE Corpus Level Views};
                \LargeTikzcirclet{A}{.025}{.973}
                \DynamicTikzcirclet{A1: \CorpusMap}{.01}{.79}
                \DynamicTikzcirclet{A2: \SearchTools}{.385}{.79}

                \draw (.155,.742) node {\LARGE Region Level Views};
                \LargeTikzcirclet{B}{.025}{.745}
                \DynamicTikzcirclet{B1: \CorpusMap}{.01}{.56}
                \DynamicTikzcirclet{B2: \RegionScatterplot}{.21}{.56}
                \DynamicTikzcirclet{B3: \RegionMat}{.455}{.56}
                \DynamicTikzcirclet{B4: \RegionList}{.685}{.56}
        
                \draw (.2,.517) node {\LARGE Neighborhood Level Views};
                \LargeTikzcirclet{C}{.025}{.52}
                \DynamicTikzcirclet{C1: \NeighborList}{.015}{.288}
                \DynamicTikzcirclet{C2: \RadialPlot}{.535}{.288}
                \DynamicTikzcirclet{C3: \NeighborMat}{.775}{.288}
        
                \draw (.173,.248) node {\LARGE Document Level Views};
                \LargeTikzcirclet{D}{.025}{.25}
                \DynamicTikzcirclet{D1: Primary Selected Document}{.01}{.0215}
                \DynamicTikzcirclet{D2: Secondary Selected Document}{.435}{.0215}
                \end{scope}
        \end{tikzpicture}
 
        \caption{
          \label{fig:views} %%\comment{Added New Figure}
          \sysname supports 
          Text Corpus Exploration 
          through a variety of flexible views which operate on multiple scales. Standard designs are enhanced with explanations and comparative features.
        }
      \end{figure*}
    }

\newcommand\introexamplefig{
\begin{figure*}
\centering
  \begin{tikzpicture}
        \node[anchor=south west,inner sep=0] (image) at (0,0) {\includegraphics[width=0.9\textwidth]{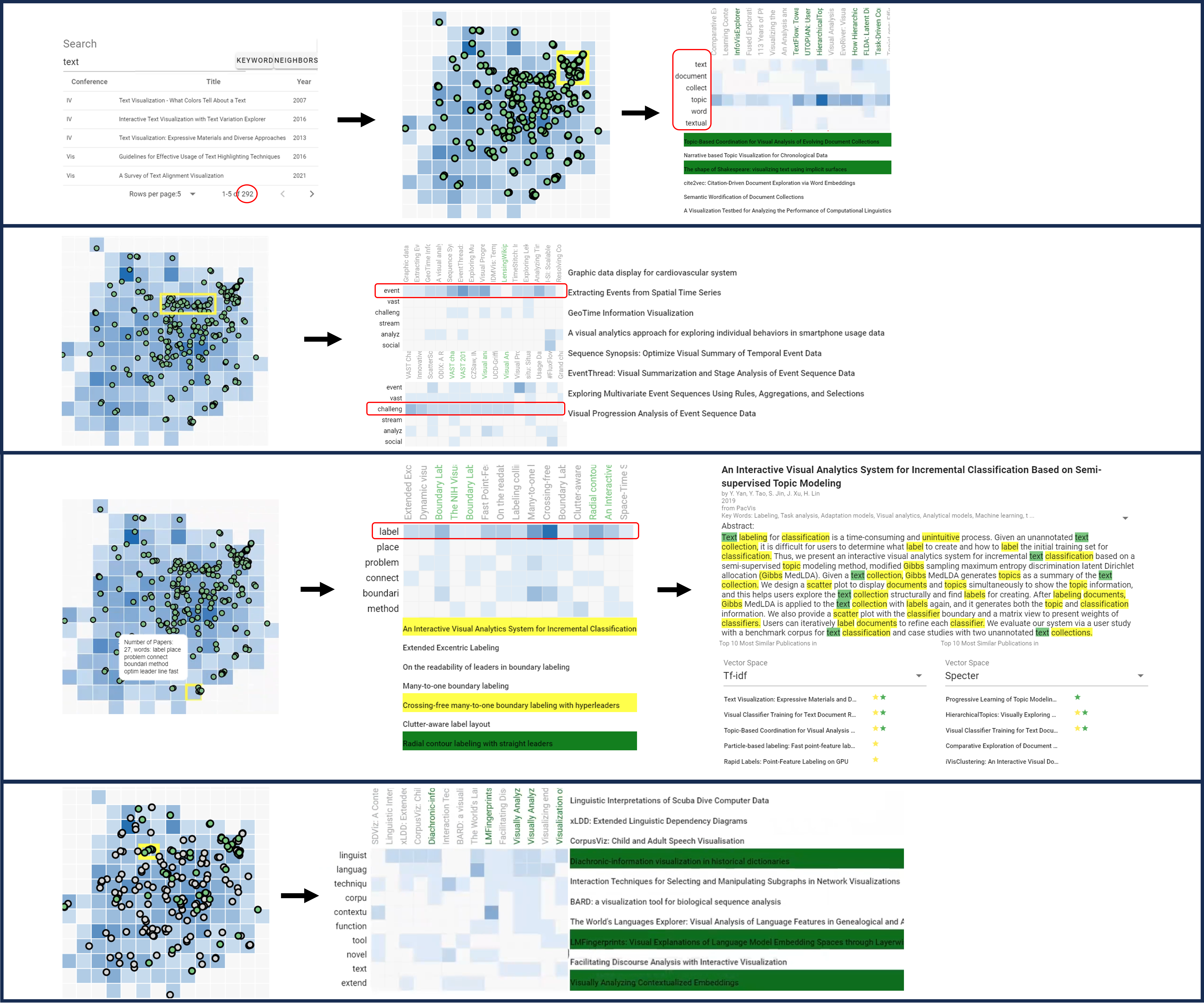}};
        \begin{scope}[x={(image.south east)},y={(image.north west)}]
           
            \LargeTikzcirclet{A}{.02}{.97}
            \Tikzcirclet{A1}{.275}{.8}
            \Tikzcirclet{A2}{.5}{.8}
            \Tikzcirclet{A3}{.74}{.8}
    
            \LargeTikzcirclet{B}{.02}{.75}
            \Tikzcirclet{B1}{.21}{.577}
            \Tikzcirclet{B2}{.65}{.577}
    
            \LargeTikzcirclet{C}{.02}{.523}
            \Tikzcirclet{C1}{.225}{.3}
            \Tikzcirclet{C2}{.52}{.27}
            \Tikzcirclet{C3}{.93}{.25}
    
            \LargeTikzcirclet{D}{.02}{.195}
            \Tikzcirclet{D1}{.21}{.027}
            \Tikzcirclet{D2}{.73}{.027}
            \end{scope}
    \end{tikzpicture}
    \vspace{.25in}
    \caption{
      \label{fig:introexamplefig} %%\comment{New Figure}
         This example illustrates four exemplary workflows in the context of the present paper. The objectives are to discover related papers (to provide context for our work and generate ideas for improvement), to learn more about the corpus and find commonly used terms. 
         \hspace{\textwidth} \\\hspace{\textwidth}  \emph{Workflow}  \circlet{A}. \circlet{A1}: We search for the relevant term \term{text}. This provides too many documents (292) to examine individually. \circlet{A2}: However, we can use the \CorpusMap to see how the documents are distributed and examine particularly dense regions. \circlet{A3}: Selecting a region (yellow rectangle) enables explanation of the region: a term-based explanation of salient words or an \replaced{exemplar-based}{item-based} explanation of representative documents. Here the terms \term{text}, \term{document}, \term{collect}, \term{topic}, \term{word} are salient and the most representative documents include other text exploration systems or have terms which suggest similar topics, such as \term{topic}, \term{theme}, \term{citation}. We identify this region as one focusing on Text Corpus Exploration, the ``TCE region:'' saving the representative document list allows for systematic, exploration.\hspace{\textwidth } 
         \\\hspace{\textwidth} 
         \emph{Workflow}  \circlet{B}. \circlet{B1}: Selecting another dense region shows with a different explanation. \circlet{B2}: The \RegionMat reveals the terms \term{event} and \term{challenge} are very salient. Sorting documents by relevant terms shows many of these papers are  Vast Challenge solutions which often involve text analysis, but are less relevant.
         \\\hspace{\textwidth} \\ \hspace{\textwidth} \emph{Workflow}  \circlet{C}. \circlet{C1}: We select a small, dense outlier region. 
         \circlet{C2}: \term{text} is not a salient term, but \term{labeling} is. \circlet{C3}: examining the selected \term{text} papers from the \RegionList, we see that many refer to text labeling, but there are text analysis systems which use labeling. While the region is generally not relevant, the specific papers can seed a similarity-based search to discover more papers about using labeling in exploration. \hspace{\textwidth} 
         \\\hspace{\textwidth} \emph{Workflow}  \circlet{D}. We want to determine which term, \term{corpus} or \term{collection} more accurately describes our work. \circlet{D1}: we search both terms and use comparison features to show both distributions on the \CorpusMap. This allows us to examine differences in how these terms are used. \term{Corpus} is localized in a few clumps (green), while \term{collection} is more scattered (gray). \circlet{D2}: Examining the clumps for \term{corpus} shows one is in a region explained by language related terms such as \term{linguist} and \term{language}, while the other is the identified TCE region. In contrast, \term{collection}'s scattered points suggest its use is more broad. Examining dense regions, we see that \term{collection} is frequently used to describe things other than text, such as images, graphs, and ensembles. The term \term{corpus} is more aligned with our usage. Other uses of \term{collection} suggest similar problems to find inspirations.
    }
    
  \end{figure*}
}
\newcommand\susstats{
  \begin{figure*}[ht]

    \centering
    \includegraphics[width=.95\textwidth]{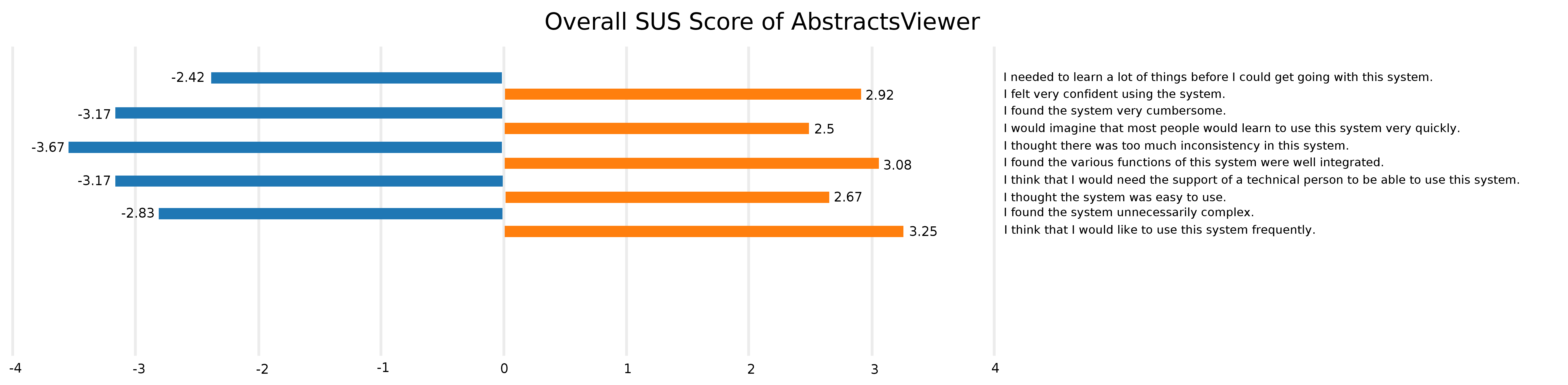};
\caption{\label{fig:sus}The SUS scores across prompts present in the user study. Numbers are in a 0-4 range and negated for negative prompts such that zero is always the worst.}
  \end{figure*}
}

\newcommand\embeddingoverlap{
\begin{figure}[ht]

  \begin{tikzpicture}

            \node[anchor=south west,inner sep=0] (image) at (0,0) {\includegraphics[width=.95\textwidth]{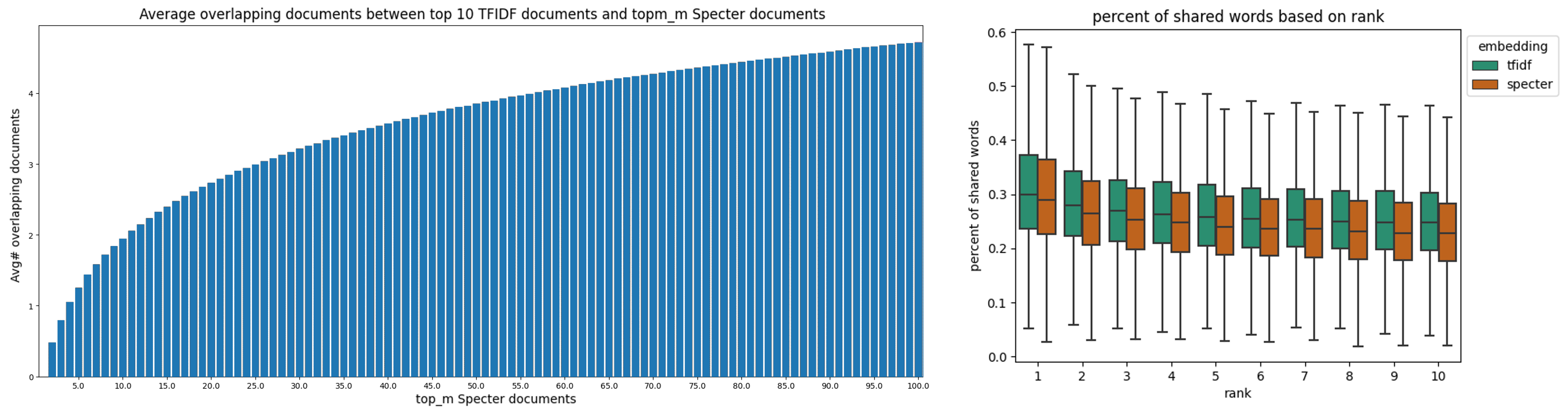}};

            \begin{scope}[x={(image.south east)},y={(image.north west)}]
           
              \LargeTikzcirclet{A}{.02}{.02}
              \LargeTikzcirclet{B}{.62}{.02}
          
              \end{scope}
      \end{tikzpicture}
      
  \caption{\label{fig:embeddingoverlap} 
  \circlet{A} The average number of documents present in both the top 10 similar documents provided by TF-IDF and the top M similar documents provided by Specter.
  \circlet{B} The average percentages of shared words between documents and their top 10 most similar neighbors, as determined by both TF-IDF and Specter embeddings. 
  }
\end{figure}
}

\onlineid{0}

\preprinttext{Authors' Version June, 2024}

\title{Enhancing Text Corpus Exploration with Post Hoc Explanations and Comparative Design}

\author{% \textit{Member, IEEE}
  \authororcid{Michael Gleicher}{0000-0003-3295-4071},
  \authororcid{Keaton Leppenan}{0009-0009-4823-458X},
  and 
  Yunyu Bai
}

\authorfooter{
  \item
  	Michael Gleicher and Keaton Leppenan are with the University of Wisconsin, Madison
  	E-mail: gleicher@cs.wisc.edu, kleppanen@wisc.edu.
  \item
  	Yunyu Bai is with Amazon.
  	E-mail: ybai45@wisc.edu.
}

\abstract{%
  Text corpus exploration (TCE) spans the range of exploratory search tasks: it goes beyond simple retrieval to include item discovery and learning about the corpus and topic. Systems support TCE with tools such as similarity-based recommendations and embedding-based spatial maps. However, these tools address specific tasks; current systems lack the flexibility to support the range of tasks encountered in practice and the iterative, multiscale, workflows users employ. In this paper, we provide methods that enhance TCE tools with post hoc explanations and multiscale, comparative designs to provide flexible support for user needs. We introduce salience functions as a mechanism to provide post hoc explanations of similarity, recommendations, and spatial placement. This post hoc strategy allows our approach to complement a variety of underlying algorithms; the salience functions provide both exemplar- and feature-based explanations at scales ranging from individual documents through to the entire corpus. These explanations are incorporated into a set of views that operate at multiple scales. The views use design elements that explicitly support comparison to enable flexible integration. Together, these form an approach that provides a flexible toolset that can address a range of tasks. We demonstrate our approach in a prototype system that enables the exploration of corpora of paper abstracts and newspaper archives. Examples illustrate how our approach enables the system to flexibly support a wide range of tasks and workflows that emerge in user scenarios. A user study confirms that researchers are able to use our system to achieve a variety of tasks.

}

\keywords{Text exploration, post hoc explanation, spatial embedding, exploratory search.}

\renewcommand{\manuscriptnotetxt}{Copyright held by the authors.}

\begin{document}

% put this here, since IEEE loads it someplace weird
% I dislike the autoref defaults
\renewcommand{\figureautorefname}{Fig.}
\renewcommand{\tableautorefname}{Table}
\renewcommand{\sectionautorefname}{Sec.}
\renewcommand{\subsectionautorefname}{Sec.}
\renewcommand{\subsubsectionautorefname}{Sec.}

%%%%%%%%%%%%%%%%%%%%%%%%%%%%%%%%%%%%%%%%%%%%%%%%%%%%%%%%%%%%%%%%
%%%%%%%%%%%%%%%%%%%%%% START OF THE PAPER %%%%%%%%%%%%%%%%%%%%%%
%%%%%%%%%%%%%%%%%%%%%%%%%%%%%%%%%%%%%%%%%%%%%%%%%%%%%%%%%%%%%%%%

% I don't know where this will go - so put it here
\newcommand{\taskfootnote}{
    \footnote{We follow \cite{rindTaskCubeThreedimensional2016} and prefer the term \emph{objective} to more specifically refer to a user's goal than the more common \emph{task}, which has many meanings.  }} % cut to fit in the literature.
% The term \emph{task} has a variety of meanings in the visualization literature \cite{rindTaskCubeThreedimensional2016}. 
% and therefore often prefer the term \emph{objective} to be clearer that we are referring to the user's goal.}
%\footnote{Following \cite{rindTaskCubeThreedimensional2016} we prefer the term objective as it more specifically refers to the user's goal than the more common term \emph{task}.}. 

\firstsection{Introduction}

\maketitle

%% \section{Introduction} %for journal use above \firstsection{..} instead
\label{sec:intro}
% Thank you ChatGPT
\emph{Text Corpus Exploration} (TCE) 
involves discovering unknown documents and gaining an understanding of the corpus, not just retrieving known documents.
As such, it shares the diverse objectives of \emph{exploratory search}\cite{WhiteBook, soufanSearchingLiteratureAnalysis2022} including to discover unknown targets and \added{to} learn about the information space (the corpus and domain from which it is drawn). This learning includes building a sense of the  documents in the collection, how they might be grouped, and how terms are used; learning is useful for further discovery and to gain understanding about the domain. 
To support the objectives of exploratory search, TCE systems extend the basic tools used for retrieval (such as keyword search) with tools based on content analysis, such as similarity-based recommendations and spatial embeddings. 
However, these tools
focus on specific objectives: they lack the flexibility to support the diverse range of tasks and workflows in exploratory search. 

% put in chat gpt version
To better support exploratory search in text corpus exploration, we propose enhancing standard tools with \posthoc explanations and comparative design.
For example, similarity search may suggest a candidate document of interest; but an explanation of \emph{why} that document is considered relevant may help a user assess the document's relevance, identify terms that could lead to better searches, or reveal interesting concepts for consideration. 
By explaining \emph{why} documents are placed in a particular spatial region of a map, a system can help a user identify whether the region is likely to be a fertile region for discovery, to see common terms that may help for search or consistency with usage in the field, or provide a set of examples of issues related to a particular topic. 
Supporting the comparison of a pair of documents can help show the relation between them for relevance checking; comparison among a group may highlight common themes.
By integrating explanations and comparisons, TCE tools can flexibly address a range of objectives and user workflows across different scales and targets.

In this paper, we examine the use of explanation and comparison as mechanisms to enhance TCE tools so they can better support users' exploratory search needs.
Our approach builds on existing tools, 
including similarity-based recommendations, embedding-based spatial maps, and term identification.
\replaced{We enhance these tools with explanations and explicit support for comparison.}
{We enhance these tools with explanations that enhance standard views. 
We also enhance standard views with explicit support for comparison, such as dual selections and juxtaposed displays.}
These enhancements enable the tools and views to address a wide range of exploratory search objectives\taskfootnote  
by integrating together in a flexible manner that allows users to dynamically improvise workflows that address complex tasks.
\added{Examples of enhancements are shown in Figure \ref{fig:teaser}.}
\teaserasfig

We introduce salience functions for ranking terms and documents as a flexible mechanism for explaining decisions made by the system including recommendations, placements, and groupings. 
These explanations are \posthoc: they \replaced{may}{do} not describe the cause of the decisions, but rather provide a plausible and interpretable reconstruction.
The \posthoc strategy 
\replaced{decouples the methods for decision and explanation, allowing us to combine}
{allows 
explanations independent of the methods used for decisions. This allows combining} state-of-art methods for similarity, recommendation, and layout with explanations.
Our salience function approach allows creating both exmplar- \added{(document)} and feature- \added{(term)} based explanations at different scales.
This allows us to use the salience functions to provide connections across a variety of views and scales.
\added{Salience ranking and highlighting also enable efficient comparison.}

\added{TCE implies exploration of a corpus, rather than a simple query or quick glance. Our approach, therefore, targets sophisticated users, such as researchers, who invest effort to understand and discover in a corpus.} 
We have applied our approach to a prototype system, \emph{\sysname} \added{(Figure \ref{fig:system}) designed to support researchers in exploring collections of short texts, such as scientific paper abstracts and newspaper leads}. \sysname combines transformer-based (and traditional) similarity scoring and recommendations, embedding-based layout, and term-based tools. It provides views for spatial and structured views at multiple scales. The use of explanations and comparison support provides connection across these views, allowing viewers to adapt them to address TCE tasks.
Our experience, including a user study, suggests that our target audience 
% \deleted{(researchers)} 
can use our approach to address a range of exploratory search objectives in text corpora.

Our overall contribution is to introduce an approach to TCE that uses explanations and comparison to enhance existing TCE tools to better support a range of user objectives. Specific contributions include: 
We introduce a \posthoc explanation approach to TCE that allows coupling transformer-based recommendations and embedding-based layouts with traditional views enabling a range of document- and term-based tasks.
We introduce a salience function approach that provides a flexible way to create explanations across scales and integrate them with views.
We present empirical evidence that \multiscale tools, \posthoc explanations, and comparative designs can provide flexibility in TCE that researchers can leverage across diverse tasks.
Our contributions are embodied in an open-source system. % available to the community.

%
%\autoref{sec:intro-example} begins with an example of our approach for concreteness, while \autoref{sec:related} puts our work in context of prior work. \autoref{sec:overview} provides an overview of our approach, \autoref{sec:details} provides details of our methods, and \autoref{sec:implementation} describes our prototype implementation and datasets. We provide evidence of our approach in \autoref{sec:eval} including more examples and a user study, and describe limitations and reflections in \autoref{sec:discuss}.
    \subsection{Overview and Example}

    % \deleted{(Old section title was: An Introductory Example)}

        \label{sec:intro-example}
        \added{
\sysname is a prototype system designed to embody our approach (\autoref{fig:system}). It was specifically designed for the exploration of scientific abstract corpora, although we also apply it to collections of newspaper articles. It intentionally focuses on content-based exploration and omits features such as meta-data analysis so we can observe how content-based tools are applied in exploration.}

\added{\sysname presents a standard TCE interface: a search panel allows for making various forms of text queries; a list view provides a structured view of query results; a scatterplot shows an embedding of the corpus with query results highlighted. However, it extends these common functions to better support the needs of exploratory search by enhancing views with explanatory and comparison features allowing for flexible, improvised workflows combining a variety of views. }

\added{\sysname also provides a set of additional views, as shown in \autoref{fig:views}, that operate at more granular scales: views of (multiple) selected documents, views of the neighborhoods (nearest documents) around these selected documents, and views of a larger, user defined region of interest. Spatial views allow for observing patterns in similarity, matrix views allow for connecting documents and terms, and list views allow for systematic examination.}

\added{The system enhances its views with 
%post hoc salience explanations and other features to enhance coordination between the views. 
explanations and comparison features.
For instance, the system uses \emph{salience functions} to identify items that can explain the sets being visualized. 
Term based salience functions select words that explain a document's inclusion in a region and list the relevant terms in a neighborhood to suggest its topic. The identified terms are seen in reorderable matrix views of document sets and text views of documents. 
Document based salience functions identify the most exemplary documents in a region providing a example document-based explanation. These views and methods are discussed more in depth in \autoref{sec:methods}.}

\added{
    \sysname provides flexibility by providing views at different scales, connected by explanations and comparisons. 
    \autoref{fig:introexamplefig} illustrates four example workflows applied in the context of researching this paper, to tasks including relevant related document discovery and assessment, identifying related concepts, and determining relevant terms. 
    The flexibility of multiple entry points for inquiry is important for discovery \cite{Thudt2012,vitality} and supported in \sysname as it allows mixing keyword search and similarity search in spatial or structured ways. Salience mechanisms provide flexible connections between types, supporting varied workflows such as outlier identification, diverse topic identification, and counterfactuals (e.g., explaining the similarity between dissimilar documents).
}

%\added{As a concrete example of these views and enhancements, we describe and illustrate an exploration of the visualization corpus related to this paper in \autoref{fig:introexamplefig} using \sysname. The four workflows detailed in \autoref{fig:introexamplefig} demonstrate how a researcher may use \sysname to discover and assess documents relevant to their work, and, in the process, explore and gain familiarity with the corpus to help them better orient themselves and position their work in the existing literature. }

%\added{In addition to those illustrated in \autoref{fig:introexamplefig}, our approach allows for numerous other workflows through the combination of different views, thereby permitting flexible exploration of the corpus. For instance, given the objective of document discovery, prior work \cite{Thudt2012,vitality} emphasizes the need for multiple entry points. As such, \sysname not only supports starting with a keyword or spatial region, as show in \ref{fig:introexamplefig}, but also with either a known document or an abstract. This enables users to select known papers or even enter an abstract from a draft of their own paper and find similar documents in the corpus. }

        %\input{sections/1-1-example.tex}
        %% note - this used to be in 1-1-example.tex
        \system
        % added intro example figure, but cannot include in change log
        \introexamplefig
        
           % added intro example figure, but cannot include in change log
        \viewsfig

\section{Related Work} \label{sec:related}
    % 

% second Docucompass paper (use for evaluation of embeddings)
% papers explictly about assessment
% Pure Suggest (beck 22)
% google paper required by reviewers

%% Recommendation Systems

Some key prior systems that inspire our work include:
Spire \cite{wiseVisualizingNonvisual} that showed the value of multiple views in TCE; 
Vitality \cite{vitality}, which showed the utility of transformer-based similarity;
Cartolable \cite{cartolabe}, which showed the value of spatial maps;
Serendip \cite{Alexander2014serendip}, which showed \multiscale tools connecting terms and documents;
and Footprints \cite{isaacsFootprintsVisualSearch2014}, which showed support for the iterative process of exploration. Our work builds on these elements, enhancing common views with explanations and comparison features.

\subsection{Text Corpora Exploration}
  \label{sec:tce-related}
\comment{Collection to Corpora}

Our goal is to support researchers in using TCE in their work.
Such work involves a variety of objectives\cite{survey-vis-sci-literature-2017,textvis-taxonomy}.
Soufan \etal \cite{soufanSearchingLiteratureAnalysis2022} document how TCE is a form of \emph{Exploratory Search} (ES). In ES, users often have diverse and uncertain goals, lack familiarity with the corpus or the domain, and lack specific search targets. 
ES is often characterized by the broad objectives of discovery and learning \cite{marchioniniExploratorySearchFinding2006}. 
\emph{Our work seeks to support the exploratory nature of TCE.}

Discovery %of documents % for dangler
without a specific target is a key element of exploratory search. 
\added{Thudt \etal}\cite{Thudt2012} describe this as serendipitous discovery (finding things without knowing what to look for), and discusses conditions that 
\replaced{promote such discovery.}{enable such discoveries by promoting encounters with items they may not have sought.}
Systems have used these concepts for exploration of text corpora \cite{vitality,Alexander2014serendip}, networks \cite{kairamRefineryVisualExploration2015}, and faceted relations \cite{dorkPivotPathsStrollingFaceted2012}.
\added{However, Andr{\'e} \etal}
\cite{andreDiscoveryNeverChance2009} explain how serendipitous discovery requires more than chance encounters; it requires understanding what is found and assessing relevance.
\emph{Our approach combines the elements of serendipity search with enhanced tools that connect to understanding and relevance checking.}

Learning objectives in TCE are less specific than discovery. To support it, tools often focus on corpus level overview. Some tools automatically
uncover structure in the documents, such as identifying clusters \cite{iVisClustering}, themes \cite{tiara,El-Assady2018}, groups \cite{Alexander2014serendip}, taxonomies \cite{BrehmerOverview}, and trends \cite{douTopicTimeOrientedVisual2016}. Such approaches directly serve tasks that involve analyzing these larger scale structures \cite{textvis-taxonomy}. However, they rely on automation to succeed, and require approaches for interpreting these often complex results \cite{chuangInterpretationTrustDesigning2012}.
Serendip \cite{Alexander2014serendip} shows the potential of bridging scales, connecting corpus scale topics through terms to specific text locations. 
\emph{Our approach uses explanation and comparison to help interpret larger structures and enable \multiscale learning; integrating learning with discovery.}

%% Switch to TCE techniques
TCE systems may either use meta-data (e.g., keywords, authorship and
\replaced{citations) or}{citations)or} content for search and organization \cite{survey-vis-sci-literature-2017}. 
\added{Meta-data approaches can serve a variety of tasks \cite{dattoloAuthoringReviewingBibliographies2022}.}
\emph{We focus on content-based TCE.}

A common content-based TCE tool for discovery is similarity-based recommendation (or search). Such tools are built into digital library interfaces (e.g., IEEE Explore \cite{IEEEXplore} and PubMed \cite{pubmed}). Services such as Semantic Scholar \cite{semanticScholar} provide more sophisticated recommendations. Vitality \cite{vitality} integrates state-of-art recommendations into a visual interface. \emph{Our work enhances this integration using explanations to allow for flexibly addressing a broader range of objectives.} 
Pure Suggest \cite{beckVisuallyExplainingPublication2022} considers a related problem explaining citation rankings.

A similarity metric drives content-based recommendations. Vitality \cite{vitality} shows the benefits of a state-of-art specialized metric in TCE. 
Many approaches exist for computing similarity, see \cite{text-similarity-survey-2013} for a survey and comparison.  \cite{witschardInteractiveOptimizationEmbeddingbased2022} combines methods and provides tools for interactively optimizing combinations. 
For graph exploration, \cite{crnovrsaninVisualRecommendationsNetwork2011} considers multiple similarity metrics, but blends them together. Accuracy in recommendations does not always correlate with user experience,  methods that can better support user tasks may be preferable \cite{Parra2014}.
\emph{Our work provides multiple metrics (including the same method as Vitality \cite{vitality}) to users in order to provide multiple perspectives. We also provide explanation tools to better integrate the recommendations into workflows.}

Automatically created 2D maps of a corpus to indicate relationships among documents were introduced
in early systems such as SPIRE (Galaxy) \cite{wiseVisualizingNonvisual} and WEBSOM \cite{websom1998}.
Such approaches embed texts in 2D based on their content and provide visualizations (often augmented scatterplots) to view this map. Recent examples include Cartolabe \cite{cartolabe}, which features high-performance, scalable scatterplots of co-embeddings of documents and other entities, 
\added{Vitality}\cite{vitality} which provides efficient scatterplot navigation on standard DR of a state-of-art metric, and Docucompass \cite{docucompass} which introduces specialized interactions for document maps.
Our work uses simpler scatterplot interactions in order to explore interpretability of the maps; our approach would benefit from the scalability concepts from these systems.
As with prior work, e.g., \cite{websom1998,nocajOrganizingSearchResults2012}, our approach uses the map to contextualize search results as well as organize information. 
\emph{Our work enhances corpus maps with explanations to better integrate with coordinated views and support a broader range of tasks.}

A wide range of layout algorithms have been used to make corpus maps.
We adopt a standard approach of using embeddings of document vectors, similar to \cite{vitality,cartolabe}. Our system does allow for switching metric and layout method, but users rarely deviate from the default. \emph{Our work mixes layout and similarity approaches to serve as explanation surrogates and provide multiple perspectives.}
Brehmer \etal \cite{BrehmerOverview} note that spatial layouts are inconvenient for some tasks, and suggest structured views (lists) to scatterplots to encourage systematic  exploration. \emph{Our work combines spatial and structured views using coordination, ranking, and salience functions to enable users to coordinate views and gain the advantages of both.}

Labels that describe the content of areas of corpus maps have been used since early systems \cite{wiseVisualizingNonvisual,hetzlerAnalysisExperiencesUsing2004}. Systems identify coherent document groups (usually by clustering) and provide labels \cite{Choo2013utopian, hanLabelTransferIntegratingStatic2018}. Systems outside of the text domain have applied region labeling to more arbitrary explanations of high-dimensional data \cite{Kandogan2012, silvaAttributebasedVisualExplanation2015}. \emph{We apply area labeling to explain arbitrary spatial regions (not necessarily coherent groups) and to decouple term selection from the map models.}

Many text visualization systems show the contributions of individual words to \replaced{the underlying models}{model decisions} (e.g., \cite{Alexander2014serendip, ConceptVector, Heimerl2012, zimmerVisualAnalysisRelationships2017}). 
\replaced[comment="address comment about Zimmer"]{
    Our \posthoc explanations approach allows us to mix-and-match salience functions to provide matrix reordering options and views even when the underlying models do not provide interpretable weights.
    \emph{Our apprach enables using popular term-document designs in more flexible ways, including more extensive reordering options, different types of explanations, and counter-factuals.}
}
{The use of re-orderable matrices to show a collection of document/term relationships is one common strategy. \emph{We make extensive use of term weight visualizations, using them with surrogate models and multiple sorting methods.}}

\subsection{Explainations and Interpretability}
  \label{sec:xai}

Explanations are a common strategy to help users interpret system behavior\cite{interpretableml,definitions-interpml-2019,Gleicher2016,barredoarrietaExplainableArtificialIntelligence2020}. \emph{Post hoc explanations} are a strategy where explanations are created after modeling and decision-making is complete \cite{molnarInterpretableMachineLearning2019}. 
Such a strategy allows decoupling the explanation process from the decision model itself, providing explanations for models that do not provide them.
\emph{Surrogate models} are a common model-agnostic strategy. An interpretable model is built to approximate the method used for decisions \cite{molnarInterpretableMachineLearning2019}. The interpretable model is used for explanations.
This approach has been called re-projection \cite{Gleicher2016} and re-representation \cite{Craven1996thesis}.
LIME \cite{Ribeiro2016} and its successors (e.g., \cite{lundbergUnifiedApproachInterpreting2017,chenGeneratingHierarchicalExplanations2020,kyubin2019}) extend this approach with \emph{local surrogates}, where different simple models are constructed for different parts of the data set. The key insight is that a simple model is unlikely to be sufficient to capture the entire behavior of a complex model, however, it may be a good enough approximation over a small portion of the data. 
\emph{Our work brings the post hoc explanation approach to TCE, applying it to interpret recommendations and layout. We introduce a local surrogate approach that applies salience functions to subsets of the data.}

In machine learning, explanations span a spectrum of scales \cite{interpretableml,molnarInterpretableMachineLearning2019}, from local (reasons for specific decisions) to global (overall properties of the model, such as generally important features). Explanations may be either item (example) based, or feature based \cite{molnarInterpretableMachineLearning2019}.
The field focuses on classification and prediction; we are unaware of it considering TCE directly.
\emph{Our work brings explanations to TCE across a range of scales by combining item (document) and feature (term) based explanations.}

%While most XAI work focuses on classification and prediction models. However,
Explanations are valuable in recommendation systems. Early work showed that explanations lead to greater acceptance \cite{herlockerExplainingCollaborativeFiltering2000,sinhaRoleTransparencyRecommender2002}
\added{and trust \cite{puTrustinspiringExplanationInterfaces2007}.}
Tintarev and Masthoff \cite{design-eval-recomm-explanations-2011} describe many goals and evaluation metrics for recommendations. The tasks are different than in TCE which can lead to different goals \cite{tranUsersAppreciateExplanations2021}, but inspire our thinking.
Zhang and Chen \cite{explainable-recomsurvey} provide a recent survey of methods for explainable recommendation. 
Standard recommendation systems are often based on other users' behaviors \replaced{and, therefore,}{, and} require different methods than the item similarity-based recommendations used in TCE.
\added{Multi-list recommendation systems have been shown to have benefits and costs \cite{jannachExploringMultiListUser2021,starkeExaminingUserEvaluation2023}.}
\emph{We apply concepts from interpretations of recommendations and predictions.}

\subsection{Exploratory Search}
\added[comment="new section from Old S3"]{
Our approach is intended to address tasks more akin to those
of \emph{exploratory search,} which is broadly defined as an information seeking problem with open ended motivations and objectives \cite{WhiteBook}. 
The objectives \replaced{are}{can be roughly} categorized as item discovery (e.g., identifying relevant items that were not previously known) and learning (about the corpus, exploration strategies, and the domain itself). 
As discussed in the exploratory search literature\cite{marchioniniExploratorySearchFinding2006,WhiteBook} the two objectives are related. Discovery is an iterative, explorative process where a user poses a ``query'' (such as terms, a document for similarity, or a region on a map) that yields a result (set of documents) which are assessed for relevance and quality, and then a new query is formulated. At each step, the user may gain knowledge (i.e., learn) about the corpus and domain in order to select the discovered items and develop further queries. However, this knowledge may serve as an outcome unto itself, beyond its use in the discovery iteration. 
}

\added{Discovery is a focus of other systems such as Vitality \cite{vitality} and Semantic Scholar \cite{semanticScholar}, and is sometimes referred to as Serendipitous Search \cite{Thudt2012}. We seek to enhance discovery by supporting assessment, query refinement, and, more generally, learning. For example, prior systems offer similarity search as a way to identify potential discovery targets; we couple similarity search with enhancements that explain the recommendations by supporting relevance assessment (e.g., by showing the document and highlighting relevant terms) and query refinement (e.g., by exposing terms and documents that may make good queries). Our examples and studies show that these enhancements are useful for aiding discovery, but also for more general objectives in learning about the corpus and domain.
Footprints \cite{isaacsFootprintsVisualSearch2014} also considers the iterative and exploratory nature of discovery, but does not consider the integration of content-based tools.}

%\section{Approach and Design} \label{sec:overview}\label{sec:design}\label{sec:approach}
% \deleted[comment="Contents of Section 3 moved elsewhere"]  
%    {\noindent\textsf\textbf{Section 3 deleted}}

%    \input{sections/3--approach.tex}

\section{Methods and Details} \label{sec:methods}\label{sec:details}
    \replaced{
    The backbone algorithms of a content-based TCE system are the similarity metric for finding documents and the layout algorithm for spatialization (\autoref{sec:similarity}).
    To these, we add the concept of salience functions that measure the importance of \replaced{exemplars}{items} and features in the outputs of these algorithms (\autoref{sec:salience}). 
    Salience functions may measure the actual strategies used by the algorithms, in which case they are causal. However, by separating algorithms and salience functions we allow for each to be chosen independently which provides flexibility in assembling workflows, including the possibility of non-causal explanations. 
    Salience functions lead to explanations by identifying items and features to be highlighted in adapted views (\autoref{sec:views}). 
    Salience functions allow us to better leverage best practices for views, for example by providing rankings for re-orderable matrices . 
    Salience functions also work well with comparison strategies, by highlighting similarities and differences as well as providing targets for comparison (\autoref{sec:comparison}).
}{
    This section provides details of our approach. 
    SEC % \autoref{sec:similarity} 
    describes the standard TCE tools we build on. 
    SEC % \autoref{sec:salience} 
    then introduces the salience functions used to provide explanations that are employed in 
    SEC %\autoref{sec:views}
    to create enhanced versions of standard views. 
    SEC % \autoref{sec:comparison} 
    provides further extensions of these views to support comparison and integration.

}

\subsection{Similarity and Layout}
    \label{sec:similarity} \label{sec:layout}
A core element of TCE systems is a \emph{similarity metric} that measures the difference between documents. Such a metric is used to provide recommendations of similar documents and to drive spatial layouts. There are a wide variety of similarity metrics, see \cite{text-similarity-survey-2013} for a survey.
Often texts are represented in some vector space, and an appropriate vector metric is used between pairs. 

Classic approaches use variants of word count vectors. Term frequency inverse document frequency (TF-IDF) \cite{tfidf} is a prevalent approach that performs well in theory and practice \cite{robertsonUnderstandingInverseDocument2004}. Our ``TF-IDF metric'' is the cosine distance applied to TF-IDF vectors created using standard stemming and stop word removal. More recent approaches, such as SPECTER\cite{specter}, apply natural language processing tools, such as transformers, to overcome drawbacks of term-based approaches. Systems such as Vitality\cite{vitality} and Semantic Scholar \cite{semanticScholar} rely on SPECTER. 

\added{\sysname provides both a transformer metric and a term-based metric.}
In practice, we observe that classic and transformer approaches often provide \emph{different} results, which are differently useful.
We do not observe a systematic preference: both were used by all participants. In our user study (\autoref{sec:userstudy}), many participants, without being told how the underlying methods worked, articulated benefits in each: TF-IDF tending to identify specific details, while SPECTER identified more \replaced{general}{specific} themes.\footnote{
% note footnote added, but can't put footnote in a change
\added{This is not surprising: TF-IDF tends to match specific words where SPECTER can match latent concepts. Examples are provided in \autoref{sec:eval}.}
}
To encourage users to employ both types of recommendations \sysname's \NeighborList shows two lists of recommendations. The supplemental material provides statistics on overlap: the two methods usually provide different results.
\replaced{
    Comparative support identifies matches between lists.
}{
    Comparative support that highlights elements identifies matches so strong that it appears in both, and aids users in appreciating the differences between lists. 
}
%\deleted{\designchoice: \sysname allows pairing two approaches: a state-of-art transformer approach and a classic term-based approach.}

% Word vector based metrics fail to compensate for synonymns (words with the same meaning), polysemy  (words with many meanings), superfluous words (e.g., the mentions of robotics in this paper), or concepts that expressed using combinations of words.

Corpus maps place each document at a position in 2D. Modern approaches use embedding techniques applied to document vectors. For example, Vitality \cite{vitality} applies UMAP to SPECTER vectors, while Cartolabe \cite{cartolabe} applies UMAP to TF-IDF vectors processed by LSA. 
% Our prototype provides a number of map layout choices as combinations of vector spaces and dimensionality reduction techniques (see supplement for a complete list and details). 
% However, in user studies, we observe that any preference was not strong enough to overcome default bias \cite{jachimowiczWhenWhyDefaults2019}. 
%In practice, the system uses a UMAP reduction (with our default parameters) of either TFIDF or SPECTER vector distance metrics. All examples described in this paper use one of those two options.
Our system uses a UMAP \added{dimensionality} reduction (with our default parameters) of either TF-IDF or SPECTER vector distance metrics. In user studies, we observe that any preference was not strong enough to overcome default bias \cite{jachimowiczWhenWhyDefaults2019}. 
Because maps are generally created once for each corpus, it is feasible for a corpus creator to manually check and tune map parameters. However, we have empirically determined a set of default parameters that are used for all maps discussed in this paper. We had experimented with providing more options to users, but chose to use two metrics and one layout algorithm.
%\deleted{\designchoice: \sysname allows selecting a state-of-art layout and similarity metric. A consistent interface is enabled by using \posthoc explanations powered by salience functions.}

%% put design choices together to be less repetitive
\added{
\designchoice: \sysname provides choices for its underlying algorithms, providing state-of-art and classical approaches. A consistent interface is enabled by using \posthoc explanations powered by salience functions. Comparative views enable workflows that combine approaches. 
}

\subsection{Salience Functions} 
    \label{sec:salience}

A \emph{salience function} is a procedure that determines an importance for each member of a set. We use salience functions for terms and documents, ordering them to provide concise lists (top-N) that serve as descriptive explanations of sets independent of how the set was determined. 
%In explanable AI (XAI) terminology, this is referred to as \emph{post hoc explanation} since it is created after the fact. 
This is a \emph{post hoc explanation} (see \autoref{sec:xai}).
\added{
    The salience functions may be chosen independently of the algorithm that determines the set; if the two do not align, the explanation will be non-causal. 
}
Such explanations are different than causal explanations: they do not try to explain the reason why the result was obtained, and may have different utility \cite{liptonContrastiveExplanation1990}.

For example, if SPECTER recommends a document as similar, the real cause involves its internal operation; a causal explanation may be useful for an AI scientist tuning the method. For a user, a \posthoc explanation that highlights a few salient words in the document may help indicate what features of that document might be relevant in quickly determining what it is about, why it is unique, or how it might have been found by a traditional search (which could help in finding more documents). Similarly, a \posthoc explanation of showing other documents that are similar might give a sense of what the document is about, why it might be interesting, and potential directions to continue exploration. These examples of feature-based (terms) and 
\replaced{exemplar-based}{exemplar or item-based} (documents) explanations may not indicate which of SPECTER's neurons are responsible, or even what words actually influenced those neurons. In fact, the explanations could be generated even if a different algorithm was used to make the recommendation. 

\added{Our approach applies salience functions to select sets, rather than the entire (large) corpus. This reduces the need for approaches that can scale to provide meaning at scale, and instead focus on methods that provide local explanations (see \autoref{sec:xai}).}

\designchoice: A key idea in our approach is that salience functions provide a flexible approach to creating post hoc explanations as they can be coupled with views that operate at the various scales of inquiry. 
\added{Salience functions are used in different views to highlight terms and documents creating explanations.}
We use term salience functions to create feature-based explanations, and item salience functions to create exemplar-based explanations. 

\subsubsection{Term Salience}
  \label{sec:term-salience}

Term salience models are used to identify and rank which words are most likely to contribute to the similarity between documents or the their inclusion in a set. 
Term salience is a variant of the keyword extraction problem. Indeed, one form of salience metric is \emph{descriptive}: scoring words based on how well they describe the set of documents. A number of established options exist \cite{hasanAutomaticKeyphraseExtraction2014}, and several are available in our system. However, for our uses, we generally prefer \emph{constrastive} metrics that find words that differentiate the set from the rest of the corpus. For this, we use the common G2 metric \cite{g2}. The G2 metric has the disadvantages that it is symmetric (it identifies words both under- and over-represented in the document set), and is purely contrastive (it doesn't consider what describes the set). To address these issues, we have created two other metrics.
The \emph{uniqueness} metric is a measure of contrast that measures how unqiue a term is to a particular region, computed as ${df}_r/{df}_g$, where $df$ is the document frequency of the term for the region ($df_r$) or the corpus ($df_g$).
The \emph{differential} metric that tries to balance description and contrast:
% df_region/R-(df_global-df_region)/(G-R)
\begin{equation}
  {m}_d = \frac{{df}_r}{n_r} - \kappa \frac{({df}_g-{df}_r)}{n_g - n_r},
\end{equation}
where $n$ is the number documents in the region ($n_r$) and corpus ($n_g$). $\kappa$ is a constant used to balance between description and contrast and is usually set to 1.
The three metrics provide options that emphasize different words. We allow the user to switch between them.

\replaced{For pairs of documents, explanations show the terms that make the documents similar.}{Explaining the similarity between a pair of documents is a special case of term salience. Often, we would like to identify what makes the pair similar.}
Salience is the amount a word contributes to the similarity metric. 
For complex metrics, such as SPECTER, feature salience approaches such as perturbation or differentiation \cite{molnarInterpretableMachineLearning2019,atanasovaDiagnosticStudyExplainability2020} could be used, in theory. In practice, we have found such approaches to be ineffective (see \cite{atanasovaDiagnosticStudyExplainability2020,xaiTransformers2022} for potential reasons).
Instead, we define a term-based salience function for document pairs based on TF-IDF: common words have non-zero salience proportional to their term frequency and inverse document frequency. This salience correlates with causal explanations of term-based metrics. In practice, it often provides useful term selection even for SPECTER. Empirically, we see similar numbers of words highlighted in SPECTER and TF-IDF matches, with the number of matches decreasing as the similarity decreases (see the supplement for statistical details).

\designchoice: \sysname uses the term salience metrics to enhance its various views (described below).
Term salience is used to identify potentially interesting words in a single document, to highlight the similarity between a pair of documents, and to identify relevant words for defining neighborhoods and regions. Salience functions highlight terms in document views, selecting terms for hovering in maps, and emphasizing terms in matrix views.

\subsubsection{Item Salience}
  \label{sec:item-salience}
Item salience models determine how relevant an item might be to the situation. The models are used to order a list of documents, enabling the user to skim more quickly. Item salience models are used to sort the search results and lists of documents in regions. 
Our system provides a number of item salience functions. The \emph{search relevance} sorting is provided by the underlying text search engine, and is only available for the text search list. The search list may also be sorted by similarity to a user selected document. 

Item salience functions are used to select exemplars from a set of documents. \emph{Similarity to the selected document} can be used to organize a set when there is a selection - whether or not that selection is in the set. Other salience functions do not require a selection. 
The \emph{neighbor list} metric approximates centrality to the set by counting the number of members of the set in each document's near neighbor list. The \emph{region words occurrence} metric orders the documents by the number of words determined to be salient in the region as a way to find the documents that best fit the description of the region provided by the salient words.

\designchoice: \sysname uses document salience functions to enhance various views so that a short list of documents can be descriptive of a larger set  \added{and such that lists can support workflows such as counterfactuals (e.g., looking for items similar to something not in the list, or similarities between items not considered similar)}.

\subsection{Views}
    \label{sec:views}
    
\replaced{
    \sysname provides a standard, multi-view, TCE interface. We have purposefully chosen familiar designs to ease learning the system. However, we enhance standard designs using salience functions for explanation, and some specific design decisions to support comparison. These functions enhance the views and help build connections between them.
    See \autoref{fig:views} for a catalog of views in our system.
}
{
    \sysname provides a number of views with multi-view coordination. We intentionally chose familiar visual representations, but enhance them using salience functions for explanation, and some specific design decisions to support comparison.
}

\added{
    \sysname provides views that operate at different scales.
    Corpus exploration involves a number of scales\cite{Alexander2014serendip}.
    We identify four distinct levels. The largest \emph{corpus} scale refers to either the entire collection, or a subset that does not have a known coherence. The \emph{region} scale refers to a subset that has a coherent thematic relationship and/or is a determined subset of interest. We use the term region for this scale because in our system, these subsets are represented as spatial regions in the 2D embedding of the corpus, although, in general this scale could extend to identified clusters or topics. The \emph{neighborhood} scale refers to a small region centered around a particular document containing its nearest (most similar) neighbors. A document can have many different kinds of similarities to other documents, therefore its neighborhood (the k-nearest neighbors in high dimensions) can be a diverse set that show multiple facets. When placed in the embedding, a document must have a single position - ideally positioning it with documents that share a common focus. The \emph{document} scale represents the smallest scale of specific, individual documents. 
}

\designchoice: 
\replaced{
    \sysname provides views for each of these scales.
}
{
    \sysname provdes a set of views at different scales to support the scales of inquiry we observe in exploration: individual document, local neighborhood (near neighbors of a specific document), region (related group), and the whole corpus.
}
Having views for different scales allows for focus+context\cite{furnasGeneralizedFisheyeViews1986} workflows and permits the choice of designs specialized for each scale. High-quality \multiscale views have been demonstrated for text exploration (e.g., the efficient scatterplot scaling of \cite{cartolabe} or \multiscale text displays of \cite{kochVarifocalReaderInDepthVisual2014,Correll2011}) and would complement the simple views in our prototype.

\designchoice: At each scale, \sysname provides both a spatial view and a structured view. Brehmer and Munzner\cite{BrehmerOverview} note that structured (list) views are preferable for methodical scanning, while others have noted that spatial views can support pattern and group finding \cite{cartolabe} or can provide context for items and terms \cite{Alexander2014serendip}. To support the widest range of tasks, we provide both types of views at each scale. \added{\autoref{fig:views} organizes the views by scale.}
Having many views has costs: users must learn about views and select between them, and screen space is required to show them. 

\deleted{See the supplement for a complete list and details of view types.}

At the corpus-level scale, The \SearchTools (\autoref{fig:system}A) allows the user to perform a search, either using terms for a full-text search, or by providing an abstract that is used for similarity search. The \SearchList provides a corpus-scale structured view of documents. It can be filtered and sorted by metadata, search results and relevance, or distance to the selected document. The \MapView (\autoref{fig:system}B) provides a spatial overview of the entire corpus. 
% The map is presented as a scatterplot, using techniques to address the scalability challenge of potentially tens of thousands of documents. Our current implementation uses a binned heatmap \cite{binned-scatterplots} of density as saturation and value  as a global context view. 
The map is presented as a scatterplot, using a binned heatmap \cite{binned-scatterplots} to provide scalability.
To provide details (i.e., Zoom), the user can select a region by clicking a heatmap cell or sweep selecting a rectangle. The \MapView combined with the \RegionScatterplot provide a focus+context view of the corpus. The \MapView shows the search results, selected items, and neighborhoods as scatterplot points overlaid on top of the heatmap. The \MapView also indicates the selected region with a yellow border. Hovering over a grid cell shows information about the cell, including the most salient terms and the number of documents in the cell.

Three views provide information about regions.
The \RegionScatterplot (\autoref{fig:system}C) provides a spatial view of the region as a Zoomed in view of the selected region of the \MapView. In the future, a continuous scaling design, as used in \cite{cartolabe} may provide better scalability. The \RegionMat (\autoref{fig:system}D) provides details about the selected region as a matrix of salient terms and documents. It effectively provides a term-based description of the region. The matrix can be reordered by choosing term- and \replaced{document-}{item-} salience functions. The \RegionList (\autoref{fig:system}E) provides an \replaced{exemplar-based}{item-based} description of the region and allows for systematic exploration. The list can be reordered by choosing an \replaced{document}{item} salience function.
\added{Selectable salience functions enhance standard re-orderable lists and matrices.}

Neighborhood views show information about the neighborhoods around the selected document (or documents if dual selections are in use). The neighborhood is defined as the $n$ closest neighbors in the selected vector space. While $n=10$ by default, it can be adjusted by the user.
The \NeighborList (\autoref{fig:system}H) shows the neighbors in rank order. The view shows a list for the selected vector space, but also a list for an alternate vector space. The coloring enables identifying common elements between lists (both for alternate spaces and between dual selected documents).
The \NeighborMat (\autoref{fig:system}F, provides a matrix view indicating the salient words within the neighborhood.
The \RadialPlot (\autoref{fig:system}I) provides a spatial representation of the neighborhood using an approximate  SolarView \cite{SolarView} radial embedding view of the neighborhood. This view preserves distance to the selected element, and shows grouping structure of the neighbors.

The \DocView (\autoref{fig:system}G) shows details of the selected documents.
It uses word highlighting to indicate salient words. Search terms are highlighted in green. 
Shades of yellow are used to indicate word salience. If a single document is selected, the G2 metric is used to differentiate the document from the corpus. If two documents are selected, the TF-IDF term salience (\autoref{sec:term-salience}) is used to show the similarities.
\added{Optionally, a second color (cyan) is used to highlight words salient in the document's region, providing an explanation of why the document appears where it does in the map.} % region words
Word highlighting allows for estimation of quantity \cite{Correll2013}, although we do not correct for word length bias \cite{Alexander2018}.
Double-clicking a highlighted word causes it to change color across the interface, assisting in identifying matches and finding the word in other views.
The \DocView allows for highlighting to be disabled if the user finds it distracting (i.e., for close reading), and provides a toggle that adds the document to the favorites list.

\subsection{Design for Comparison and Linking}
    \label{sec:comparison}

\sysname uses selection and highlighting to build links across views. It has a notion of a selected document and a selected region (which may not contain the document). Regions may be specified by either sweep selecting in the map view, or clicking on a heatmap grid cell. The selected document may be specified throughout the interface: picking from a list or clicking an item in a scatterplot or matrix. 

To better support comparison, \sysname has a notion of dual selection: the user can select two documents. 
Dual selections are shown in a similar manner. Both  define neighborhoods, and both neighborhoods are shown across the interface.
The second selection allows for comparison in several ways: the two documents appear side-by-side in the \DocView, with similarities highlighted; the neighborhood lists can be compared, aided by coloring;
the neighbors are shown in different colors in the various spatial views, which can help contrast the contents. Additionally, the \NeighborMat shows both neighborhoods for comparison.

\sysname uses a consistent coloring scheme for selections. The selected items are shown in yellow, neighbors of the first selection are shown in orange, and neighbors of the second selection are shown in pink. Green is used to indicate search results. These colors are used in the scatterplots for points (the disc is split with each side corresponding to one selection). 
\added{The \MapView can show the previous search result in an alternate color to help comparison.}
The colors are used in list views (including the document titles in matrices), except that the search list does not show green (since all items would have this property). They are also used in the matrix views to color document titles and the radial views to color the outline of points.

% When coloring, selection has precedence over the first neighborhood (orange) which has precedence of the second (pink).

\sysname also provides workflow support including a history to allow returning to a previous point in the exploration, a shopping-cart style favorites list to record discovered documents, linking to digital libraries, and exporting of the discovered lists.

\section{Implementation and Data}\label{sec:implementation}
    % \subsection{Implementation}
\replaced{
    % old 3.1 text
    \sysname is a prototype system designed to embody our approach. It was  designed for the exploration of scientific abstract corpora, although we also apply it to collections of newspaper articles. It intentionally focuses on content-based exploration, omitting features such as  meta-data analysis so we can observe how content-based tools are applied in exploration.
}
{
    The \sysname prototype system implements our approach. 
}
The system has four components: a web-based front end that runs in browser, a server-based back end, a tool for preprocessing corpora to be used with the back end, and tools for scraping digital library web pages to construct corpora.

\noindent\textbf{Front end:} The front end is written in JavaScript using the Vue\footnote{\added{\url{https://vuejs.org/}}} framework and the D3\footnote{\added{\url{https://d3js.org/}}} visualization library. % It is served as a static web page and connects to the backend.

\noindent\textbf{Back end:} The back end is written in Python. The back end loads precomputed information about the corpus at start time. It is run with a single specific corpus; to support multiple corpora, we create multiple instances of the back end that appear as different web servers. For the experiments described in this paper, a back end instance for each corpus is deployed as a container on a cluster-based VM that runs on a departmental server. The VM is configured with VMware ESXi to have 4 virtual CPUs and 8GB of RAM and uses Portainer to run multiple containers.
The back end uses the Whoosh \cite{whoosh} library for  indexing and search. It provides a query language, and handles word variants.

\noindent\textbf{Precomputation:} A Python-based application converts corpus data from standard bibliography files into data for the back end. The system precomputes the vectors for the various metrics, the embeddings for the \MapView, each document's 100 nearest neighbors, and word stems.
The preprocessor uses standard Python libraries (SciKit-Learn \cite{pedregosa2011scikit} and NLTK \cite{nltk}) for text processing, constructing embeddings, and precomputing nearest-neighbor lists.
Text processing is done using the default parameters for NLTK, and includes stemming and stop word removal. These processed texts are used for vectorization and salience analyses.
Vector spaces are computed with standard Python \replaced{libraries}{libraies}. 
\deleted{Details of the various vector space computations and salience metrics are provided in a supplement.}

% For vectorization metric, the system includes TF-IDF, , Non-negative Matrix Factorization (NMF) \cite{fevotteAlgorithmsNonnegativeMatrix2011} and Specter .
%The NMF model was built with $n_components$ set to 10 to get 10-dimensional vectors, $random_state$ set to 1 for reproducible results,
%the regularization term alpha set to 0.1, and $l1_ratio$ set to 0.5 for the combination of L1 and L2 penalty.
%For \MapView's embeddings, the system includes UMAP \cite{umap}, LSA \cite{lsa}, T-SNE \cite{tsne}, MDS and non-MDS \cite{Carroll1980}.
%We set UMAP's $n_neighbor$ to be 10 because we are showing the top 10 recommendations by default. We used cosine metric to preserve the distance in high-dimensional space.
%The dissimilarity parameter for MDS and non-MDS is a precomputed distance matrix using $pairwise_distances$ with cosine metric.

\noindent\textbf{Data Scraping:} Like Vitality \cite{vitality}, we have built Python tools for retrieving bibliographic data from digital libraries. Our tools automate collecting abstracts for conferences and journals, and create visual summaries of corpus statistics so users can check coverage.

\noindent\textbf{Corpora:} For our initial usage and testing we have developed several different corpora to use with \sysname. Our data is based on the abstracts (or leads from Newspaper articles).
% Creating similarity metrics and embeddings based on the full-texts is possible, but can lead to different results \cite{compare-full-text-abstracts,embComp}. For each corpus, we provide both the number of documents and the number of terms (distinct stems, after removing stop words) as system performance depends on both factors. 
The corpora used in our experiments are:

\begin{enumerate}
    \item \emph{Visualization:} A \replaced{corpus}{collection} of paper abstracts from a variety of visualization venues. 
    %We include journal special issues for the IEEE Vis and EuroVis conferences, but not the regular issues of the broader journals they appear in. We include key visualization focused journals (e.g., Information Visualization), but not more general conferences (such as CHI). 
    \deleted{The corpus comprises} % keep consistent
    (5237 abstracts and 15600 terms)
    \item \emph{Robotics:} A \replaced{corpus}{collection} of abstracts from a variety of papers from robotics and haptics venues. %This collection includes many major and minor robotics conferences and journals. The corpus has 
    (42114 abstracts with 52640 terms)
    \item \emph{Recent Robotics:} A subset of the Robotics corpus comprising papers published after 2016. 
    (24785 abstracts and 39034 \added{terms}) 
    \item \emph{NY Times 60K:} A \replaced{corpus}{collection} of newspaper article leads as described below. 
    (60000 articles and 109580 terms)
\end{enumerate}

To provide testing corpora of different sizes with similar content, we created corpora by downsampling the \textit{The New York Times Annotated Corpus} \cite{sandhaus2008new}. We use the ``lead'' from each article (generally the first paragraph). Our ``60K'' data set consists of 250 articles randomly selected from each month from 1987 to 2006. Our prototype was able to operate on a 120K \replaced{corpus}{collection} (500 articles per month), but was ineffective \added{(see \autoref{sec:limitations})}.

\noindent{\emph{Performance:}} Our not-well-optimized prototype provides reasonable performance on the corpora we described. Sweep selection of regions in larger corpora is problematic because it requires computing region term statistics; for fixed (grid cell) regions, salient terms are precomputed (to enable the hover lists over the \MapView).
On the smaller data sets, performance is much better. Scalability is a limitation of our prototype that we discuss in \S\ref{sec:limitations}.

\section{Assessment}\label{sec:eval}
    We provide evidence of the utility of \replaced{our explanation and comparison enhancements}{our approach} through both a set of usage examples, and controlled user studies to see how researchers employ our tools. 

\subsection{An Example Exploration} \label{sec:awareness}

    \awarenessimage

\added{This example considers an exploratory search to identify entry points into the literature. It shows how explanation and comparison features help in the common pattern of refining a vague concept to reveal how papers describe a problem (terms), as well as to find some initial papers to learn about what has been done already (and provide for further citation-based gathering).
The topic is enabling human awareness of robots in human-robot collaboration: how can people be informed about what the robots in their environment are doing? 
No obvious keyword search was specific enough to expose good entry points - the keywords are common, and generally lead to large numbers of papers about different problems (such as robots being aware of people).
}
%\deleted{
%        We were beginning a new project about enabling human awareness of robots in human-robot collaboration, that is, how can people be informed about what the robots in their environment are doing. 
%        We sought entry points into the robotics literature: to understand how roboticists describe this kind of problem (terms), as well as to find some initial papers to learn about what has been done already (and provide for further citation-based gathering). 
%        Unfortunately, no obvious keyword search was specific enough to turn up an entry point - the keywords are common, and generally lead to large numbers of papers about different problems (such as robots being aware of people). 
%        This example shows a standard exploratory search pattern in our system: a broad keyword search is narrowed by identifying regions on the map with explanations, and then  term and item-based explanations lead to improved queries.}

We begin with a keyword search for the term \term{awareness} (stem \term{aware}). This returns 500 hits in the Robotics 2016-2022 corpus \autoref{fig:awareness}(left). We observe many clumps of papers in the map, and begin a process of manually selecting regions to see if they may be of interest. The selected regions contain more than the search hits and provide context to the results. In some cases, the theme of a region is clear from its salient words, but it was unclear why awareness comes up. For example, a region had terms that implied it was about depth estimation; sorting its documents by 
% \deleted{saliency of} % dangler
search terms revealed that awareness was used to refer to methods (e.g. ``geometry-aware''). In other regions, the salient terms were unhelpful, but the exemplary documents revealed a clear theme. For example, a region that had salient words relating to broad concepts \term{simulation} and \term{trajectory} could be seen to be about autonomous vehicles. 
\replaced
    {Explainations enabled quick dismissal of these regions.}
    {These regions were easy to dismiss as unlikely to be fruitful locations to explore.}

\replaced
    {In another area, explanations help identify a promising region by highlighting salient stems (\term{human}, \term{collabor}).}
    {A region had promising salient stems (\term{human}, \term{collabor}).}
\replaced % get rid of danglers
    {The salient (exemplary) papers are}
    {Looking at the salient (exemplary) papers showed papers}
related to human performance in human robot collaboration, a promising area. Re-sorting the documents by search terms reveals a hit about ``operator awareness'' \autoref{fig:awareness}(left). % vibrotactile
The paper itself had irrelevant neighbors that shared methodological details \autoref{fig:awareness}(center), however the phrase seemed appropriate. The exact phrase does not occur elsewhere in the corpus, but searching for \term{operator} and \term{awareness} revealed a list of promising papers. One of the first hits seemed particularly useful, the title described an empirical study of effects on various aspects of operator performance, including awareness. 

\replaced
    {Using this relevant paper as an anchor for similarity search (i.e., the strategy shared with prior systems) was unfruitful.}
    {The paper was relevant because it showed an empirical approach to understanding human awareness of robots. However, using it as a starting point for similarity search was unfruitful.} 
\replaced
    {Our approach allows us to understand why: using}
    {Using}
 the \NeighborMat shows that its neighbors relate to different aspects of the paper \autoref{fig:awareness}(right). 
\replaced{Our approach also allows us use the paper as an examplar of what we are looking for in a region more likely to contain relevant papers.}
    {However, we could still use this paper as an exemplar of what we are looking for (empirical studies of operator performance).} 
\replaced
    {Similarity to this paper serves as a salience function to identify appropriate papers in the \term{awareness} region identified above.}
    {We used similarity to this paper to sort the \term{awareness} papers, bringing those interested in human performance to the top of the list.}

\subsection{\replaced{Usage Vignettes}{Additional Examples}} \label{sec:examples}

\replaced
    {We provide a series of short vignettes that exemplify how elements of our approach enable users to achieve their objectives. Each vignette is chosen from a more complete use case to highlight a specific aspect of the system.}
    {These examples are from actual use of our system and the unstructured explorations of study participants.} 
In some cases, we recreated the examples with different topics to preserve confidentiality of research in progress. 
% \deleted{The supplement provides more details and examples.}

    %%% Relaxed IK Example
\noindent\emph{Example: multiple similarity metrics and \posthoc explanations:}
\replaced
    {A roboticist works with a specific algorithm (\emph{RelaxedIK}) that addresses a standard robotics problem (inverse kinematics, IK). They are}
    {A robotics laboratory makes extensive use of \emph{RelaxedIK}, an algorithm for solving a common class of robotics problems (inverse kinematics, IK). A researcher is }
interested in finding papers that provide competitive methods and potential new applications. 
\added{By using different similarity metrics, they can address both questions, and by using \posthoc explanations they can see how the metrics differ.}

\replaced %dangler
    {They}
    {In \sysname, they} 
search for and select the RelaxedIK paper. The neighbors for the TF-IDF metric form an extremely tight cluster 
(all the scatterplot points are so close that they are covered by the star glyph in the map requiring the \RegionScatterplot to see them)
in the \MapView with one outlier. 
The outlier is a paper that uses RelaxedIK.
Examining the neighborhood in the \NeighborMat shows that all papers use similar methods for the same problem (constraint optimization approaches to RelaxedIK). 
% terrible sentence
%Examining the region that the neighbors are contained in using the \RegionMat shows the salient terms include the method (\term{constraint}, \term{optim}, \term{solve}) but also related problems (\term{trajectory}, \term{control}).
They examinined the region that the neighbors are contained in using the \RegionMat which showed that the salient terms include the method (\term{constraint}, \term{optim}, \term{solve}) but also related problems (\term{trajectory}, \term{control}).
Sorting the \RegionList by centrality reveals that the exemplary papers all address various robotics problems with constrained optimization solvers. Sorting the \RegionMat shows all of these documents (including RelaxedIK) share terms about the method, but not the problems. 

In contrast, the neighborhood formed by the SPECTER metric has documents spread over the map with less obvious connections. The \NeighborMat with a contrastive metric shows terms \term{geometric}, \term{feasible}, and \term{end-effector}, properties useful in robotics problems (and provided by RelaxedIK). Hovering over regions for the neighbors reveals these regions are about different applications (such as grasping, tele-operation, and deformable contact), suggesting potential applications that benefit from RelaxedIK's properties.
\added{The different metrics provided differently relevant sets of papers, while the explanation visualizations allowed for rapid interpretation of these suggestions.}

%%% Carter's Device
\noindent\emph{Example: term refinement and related problem identification:} A robotics researcher was applying a new computer vision device for robotics application. They were trying to identify similar devices, the methods used with them, and their applications. One lead came when a review of a manuscript provided a related paper. Examining the neighbors of the paper in the \CorpusMap showed them to be in regions defined by common robotics problems (e.g., \term{slam}, \term{localization}) and methods applied to them (e.g., \term{trifocal}). This suggested techniques for comparison and potential applications. The \NeighborMat revealed \term{laser} and \term{rangefinder} as salient terms: looking in \DocView showed that \term{laser rangefinder} was historically a commonly used device with similar properties to the sensors being considered. 
\replaced
    {Explanations allowed similarity search to reveal better terms for subsequent searches and reference.}
    {This led to a new search that revealed relevant related work.}

%%% Yea-Seul fact vs. insight
\noindent\emph{Example: connecting terms and regions with explanations:}
%\particip{4} 
A visualization researcher sought to use a known relevant paper to find other works related to her topic and to identify the common ways authors referred to the concepts. 
To prioritize the most relevant neighbors, she used \NeighborMat to identify salient words.
She observed the term \term{fact} and used the prevalence of this term as a salience function to identify relevant documents that referred to her concept of interest. 
Within this \replaced{document-based}{item-based} description of similar documents, she observed that the term \term{insight} was often used in a similar manner, providing for both future searches as well as an alternate term for use in describing her work.
\added{This example shows how explanations allow similarity search to reveal terms.}

%%% Pragathi - projectors
\noindent\emph{Example: identifying unexpected related topics with explanations:} %\particip{5} 
a researcher sought to identify work related to a paper with a draft abstract about robot camera systems.
The neighbor search using SPECTER provided a set of neighbors that seemed irrelevant. Examining the \NeighborMat revealed the stem \term{project} as salient to the neighborhood. 
\replaced
    {This term seemed irrelevant, but examination the items using it}
    {Examination of the documents}
showed that they involved controlling projectors, which is a related problem to cameras, 
making a connection in terms of topic as well as showing a set of papers to be relevant.
\added{This example shows how explanations turn unexpected terms into concepts for exploration by connecting term- and \replaced{document-based}{item-based} explanations.}

% Bolun
%\noindent\emph{Example: gaining familiarity and ideas:} 
%to gain familiarity and trust with the system and corpus, a researcher used a keyword search for a specific topic they work on. They found the expected documents, and learned the relevant region of the map. However, looking at the most salient terms, they saw were surprised to see the prevalence of a seemingly independent concept - which motivated them to explore the connection. 
% can't really say \term{``haptic realism''} passive/stability
% 

% Why here not there?

\subsection{User Studies} \label{sec:userstudy}
    The development of our approach included use by our collaborators in their research interleaved in the development process.
Additionally, we conducted two more formal studies. The first was a pilot study ($n=6$) with an early prototype to understand how untrained participants would appreciate it. 
\added{The pilot study followed a similar protocol to the study described below, however it used an earlier prototype system with fewer structured tasks.}
The pilot study showed that researchers could apply our approach. They benefited from the enhancements we provided to the standard tools to better use recommendations and the spatial map to identify relevant documents and terms and to rapidly assess relevance.
However, it also exposed needs that led to the current designed
\replaced{For example, the need to better organize screen space and to allow for hiding seldom used features to avoid distraction.}{.} 
%For example, the pilot participants needed the structured views, 
%emphasized needs including the need to combine item- and term-based explanations; list views that support systematic exploration coupled to the spatial views; multiple, high-quality salience mechanisms to help identify terms and items of interest; effective use of screen space to show multiple views; and the need to hide seldom used features to avoid distraction. These factors led to our current design.

We conducted a more thorough user study with a later prototype. 
We sought to assess how well our target audience can use our approach to accomplish TCE tasks and observe the ways that they explore. 
To make these observations, we designed a human subjects study where participants were trained to use the system, completed a set of structured tasks that exposed them to 
% \deleted{various} % dangler
elements of our approach, and performed free exploration using the system on a topic of their own choice.

Participants needed to be researchers 
\replaced %dangler
    {(our audience)}
    {(i.e., our target audience),}
familiar with the literature in either robotics or visualization 
\replaced %dangler
    {(the available corpora)}
    {(as these were the available corpora), }
and have a specific topic to explore (we invited participants to bring an abstract of a work in progress). Each participant worked with either the robotics or visualization corpus based on their background, we developed a set of structured tasks based on each corpus.

\noindent\emph{Participants:} Our participant pool was limited because of the research experience requirements and the requirement to be available for an in-person study. Therefore, we recruited participants by convenience sampling from our colleagues working on topics related to visualization, robotics and haptics. These participants may be biased as they are all colleagues, or students of colleagues.

We recruited 12 participants with various levels of research experience, including two professors, one post-doc, six senior graduate students and three new graduate students. Four identified as female and eight identified as male. Participants' academic departments were Computer Sciences, Mechanical Engineering and Psychology. The participants were not compensated, but were eager to explore the literature related to their work.

\subsubsection{Study Procedure:} % Technical details.
The sessions took place in our laboratory. Participants used a desktop computer with sufficient performance to work well on the corpora and a 27-inch monitor.
Sessions were about 60 minutes. After providing informed consent, we began with a brief walkthrough of different views in the system. During this process, we presented specific examples of each view. Participants were then given a set of structured tasks to complete. In the final phase, we invited participants to explore the corpus freely based on their interests. At the end of the session, participants completed a demographic survey and a system usability scale (SUS) survey \cite{brookeSUSRetrospective2013}.

The sessions were recorded with both audio and screen recording. The experimenter was available during the entire session to answer questions and to encourage the participant to ``think aloud'' and explain what they were doing.

Participants completed a set of structured tasks where we asked them to answer specific questions using the corpus.
The first type of task required participants to interpret specified regions of the corpus in the map. Participants were allowed to use any views they thought were appropriate. A second type of task involved presenting participants with a paper and its neighbors and asking them to assess their similarity and relevance. The final type of task asked the participant to search for a specified term and describe the distribution of results. \added{Each task type was repeated multiple times.}

For the free exploration, participants were asked to bring an abstract of a work in progress to enable interest-driven explorations. Participants were free to discover any unexpected related work, alternative search terms or build a new understanding of the corpus.

\subsubsection{Study Findings}

\noindent\emph{Structured Tasks:} All participants were able to complete all \added{of the structured} tasks.
\added{The participants made use of all of the key features of our approach in order to complete the structured tasks, showing their utility for (at least) these tasks.}

All participants were able to explain the similarities between pairs of documents. 10 participants explicitly said that word highlighting in \DocView helped.
Multiple participants agreed that the highlights reduced time to understand the similarity. For example, \particip{5} said \pquote{Just reading around the highlights helped me get the context quickly, rather than read through the whole thing}.
However, \particip{7} acknowledged the highlights \pquote{may be distracting in reading, but helps build the connection between 2 papers}. One participant used the highlighting for seeing the structure to speed her reading. We note that the experiment required all participants to use \DocView's TF-IDF explanations to explain both TF-IDF and SPECTER similarities. All participants were able to efficiently assess relevance. Only 5 explicitly mentioned \DocView highlighting. %, others relied on close reading.

All participants were able to interpret regions using both \replaced{document-}{item-} and term-based descriptions (\RegionMat and \RegionList). All participants made interpretations that required combining different views.
They expressed a range of preferences: 7 participants preferred the simplicity of the \RegionList, 1 preferred the intuition of the \RegionMat, and 4 preferred the combination of both.

During these structured explorations of regions, participants used the multiple term salience metrics in the \RegionMat. Several identified the pattern that the differential metric brings general concepts to the top of list, while contrastive metrics bring out specific techniques. 
%For example, in the robotics corpus, a region with salient terms \term{path} and \term{plan} under the former, had specific algorithms for path planning. 
For example, \particip{5} said \pquote{I think there is more information about the specific algorithms} (with contrastive).

Participants were % explicitly -cut for dangler
asked to look at both SPECTER and TF-IDF neighbors in the process of exploration.
We did not observe a systematic preference: both were used by all participants. Many participants saw benefits in each, they observed that TF-IDF tended to identify specific details, while SPECTER identified general themes. 
Participants found value in seemingly ``wrong'' recommendations. For example \particip{12} said \pquote{So this is interesting because this isn’t directly relevant to the paper at all, but it’s like a whole other area of work that I do}.
% For example, 
% 12 “So this is interesting because this isn’t directly relevant to the paper at all, but it’s like a whole other area of work that I do”
% 12 “This highlights the global similarities (between this paper and) my research program and this (paper) is a different aspect of it. That’s pretty neat.”
% 11 "Even though it’s not really quite related specific to my paper, but these are my interest"

% At least half came up with a model of how to use each and articulated it without being prompted.

\noindent\emph{Quantitative SUS Results:} 
%The complete numerical SUS results are provided in the supplement. The results are generally favorable, but expected patterns emerge: participants found the system useful, but difficult to learn. 
Detailed SUS results are in the supplement. The results are generally favorable, but expected patterns emerge: participants found the system useful, but difficult to learn. 

\noindent\emph{Unstructured Exploration:}
All participants were able to discover relevant unexpected documents during the free exploration, which included identifying candidates and assessing their relevance. All participants self-reported that they had found something they were happy with.

In the unstructured explorations, not all participants had the opportunity to use all features (possibly because of limited time). 
For example, 8 out of 12 participants identified a neighborhood in the \MapView and explored it to identify documents of interest. Some participants only used the \MapView to provide context for documents they identified other ways.
Only two chose to  investigate outliers in the \MapView. For some, outliers were not present in their explorations, others were more focused on clusters.

The tasks in the structured exploration focused on our motivating uses: discovering candidate documents, rapidly assessing relevance, finding terms for search iteration, and identifying relevant map regions. However, observing unstructured explorations we saw a broader range of exploratory search behavior, including interesting ways to use the tools of our system to support it. 
%For example, one participant used salient word highlighting to see where documents used their key words in an effort to appreciate their structure. Another participant used search terms to highlight differences when comparing documents side-by-side. The use of the \NeighborMat to compare neighborhoods of different documents was suggested by a participant. 
We observed participants discovering far more than documents (and terms to use in searching): they found topics and themes they were unaware of, unexpected groups in the literature (e.g., \pquote{I never expected so many papers about...}), and connections that surprised them. 

Even in the short sessions of the controlled studies, we observed emergent behaviors and a variety of exploratory goals beyond our prompts. While hard to quantify, we view it as reassuring evidence that the flexibility of our approach achieves its goals. In the first hour of working with the system, users do not have sufficient opportunity to learn to work with all of the tools it provides. We noted a pattern that many participants sought to gain familiarity with the system and corpus by looking for familiar terms and documents. In the process, they both built trust by seeing expected items and often discovered things.
In less structured observations with longer-term users, we (anecdotally) observe more advanced usages.

\section{Discussion and Conclusion}\label{sec:discuss}
    %% notes

Experience with our prototype suggests that our target audience of researchers can apply our prototype system for real applications. Users interleave a variety of exploratory search objectives while working with a corpus. They discover documents by iteratively posing queries, assessing the relevance of the outcomes and learning about the corpus to improve their search, while also learning about the domain, the terms used, and the corpus. They use our variety of views, applying standard TCE tools with enhancements. They use post hoc explanations and comparison features to mix views at different scales.

However, our initial experience with the prototype also exposes limitations such as: \label{sec:limitations}

\noindent{\emph{Evaluation:}} The success of target users applying our methods in TCE provides an initial validation of our approach. However, more detailed evaluation will be valuable in understanding the trade-offs between system complexity, flexibility and automation.

\noindent{\emph{Usability:}} Our approach is based on providing users with flexibility. However, this requires them to make many choices. 
Strategies to combat the complexity include reasonable defaults, automated guidance (e.g., assistance in selecting options to achieve tasks), and culling seldom used options.

\added{\noindent{\emph{Workflow Support:}} \sysname does not provide explicit support for important aspects of workflows. For example, identified regions and groupings must be remembered by the user. Prior systems, such as Serendip \cite{Alexander2014serendip} show how this can be supported with persistent annotations. Similarly, comparisons between regions rely on manual temporal juxtaposition. Systems such as EmbComp \cite{embcomp} provide examples of how this can be better supported.}

\noindent{\emph{Metadata:}} \sysname focuses content-based tools to better explore their issues. Integrating with meta-data and citation-based exploration would be a valuable extension.

\noindent{\emph{Implementation:}} \sysname is a prototype that may not scale to a robust deployment for many simultaneous users.

\noindent{\emph{Scalability:}} Our system performs well on corpora of tens of thousands of short documents on modest hardware. However, larger corpora, strain the system in terms of performance (speed) and usability.  Recent systems, such as Cartolabe \cite{cartolabe} and Vitality \cite{vitality} provide examples of how TCE systems can be made to perform on larger corpora. Improved views will be required to operate at larger scales; lists, matrices and scatterplots become unusable with hundreds or thousands of documents in a region. Handling longer documents will require improved methods for showing similarity and salience words.

\noindent{\emph{Automated Analysis:}} \sysname requires manual identification of summary structures such as clusters and topics. Integrating automated tools will raise new usability and interpretability challenges.

%\noindent{\emph{Loop Closure:}} Our system enables the user to identify reasons why the similarity metric provides good or bad results, but provides no mechanism to use these insights to improve the metric. For example, bad matches often identify words that could be ignored, however, the system provides no ability to recompute the metric with an updated vocabulary.

\noindent{\emph{Reliability of Post Hoc Strategies:}} Post hoc explanations rely on the existence of, and the ability of algorithms to find, interpretable explanations of potentially complex things. There may be no simple explanation for why two documents are considered similar, or the surface level similarities may not be meaningful. 
In practice, we believe the post hoc strategy often provides utility, 
future studies should assess whether poor explanations are a distraction.

\noindent{\emph{Limited flexibility of regions:}} The ``regional scale'' (a group of documents that is thought to be related) is valuable in exploration, but has limited implementation in our system. 
\replaced%danglers
    {We seek to provide better interfaces for manually defining regions (both geometrically and as sets) and to integrate automation for region finding.}
    {In the future we will seek to provide convenient interfaces for manually defining regions (both geometrically and as sets) and uses this as an opportunity for integration with automation.}

\added{
    \noindent{\emph{Impact of multiple recommendation lists:}} The use of multi-list recommendations is known to have benefits and costs \cite{jannachExploringMultiListUser2021,starkeExaminingUserEvaluation2023} including positive and negative effects on trust \cite{puTrustinspiringExplanationInterfaces2007}. In the future, we will seek to balance the diversity and contrast with the extra user effort.
}

\noindent\textbf{Reflections:}
In this paper, we have explored the use of post hoc explanations and comparison to enhance existing tools for text corpus exploration. While our approach has been prototyped in the \sysname system, we note some general themes that extend beyond our system and possibly beyond text applications. The varied nature of exploratory search suggests that users need to be supported in the iterative process of document discovery (including assessment and query refinement), but they also need support for the learning aspects of exploration. This variety of objectives occurs at a variety of scales, suggesting that supporting the \multiscale nature of exploration is important. While different tools and views operate at specific scales, explanations and comparisons provide ways to link these scales. While TCE tools, such as recommendations and maps, provide value on their own, their value is enhanced by providing post hoc explanations that help connect across scales and between items and terms. These post hoc explanations can be created in a flexible, \multiscale manner by defining salience functions that can be applied to different sets and used by different views.

Initially, we attempted to define a concrete list of specific tasks, and to \added{create a} design to support these tasks. However, in practice we observed a much broader and unexpected range of user objectives and strategies, as the exploratory search literature might have predicted. We believe (and our study suggests) that our system supports key specific objectives, such as enabling document identification and assessment or term refinement. However, we also believe that the flexibility to adapt to the broad range of user objectives and workflows is a more important benefit of our approach, and is evident in our examples and study.

\noindent\textbf{Conclusion:}
Our work has explored how a TCE system can support a range of exploratory objectives in text corpus exploration. By enhancing maps and recommendations with explanations and comparisons, these tools can connect across scales and between documents and terms. We introduced salience functions as an approach to flexible post hoc explanations that work across scales and item types. We used these ideas to enhance standard views so they better support exploratory tasks and to integrate among views at different scales. Experience with a prototype implementation suggests that users can employ this flexibility to achieve a variety of exploratory goals.

\acknowledgments{This work was supported in part by NSF Award 2007436.}

% \newpage
%\bibliographystyle{abbrv-doi-hyperref}

\bibliographystyle{IEEEtran}

\bibliography{Bibs/NonZotero22,Bibs/AbstractsPaperRefs22,Bibs/AbstractsPaperRefs23,Bibs/NonZotero23}

% \includepdf[pages={1}]{biography.pdf}

\onecolumn
\appendix % You can use the `hideappendix` class option to skip everything after \appendix

\section{Quantitative Details} \label{sec:quantitative}
  
    This appendix provides quantitative details to support points in the paper.

\subsection{Usability Assessment}

\sysname is a prototype meant to explore the our approach of \posthoc explanations and comparison. As a prototype, and as a tool targeting expert users, a polished interface was not a primary concern. However, we did assess system usability through a system-level evaluation. 

A System Usability Scale (SUS) survey is a standard 
approach to system evaluation %, asking users to indicate their degree of agreement 
with 10 standardized questions \cite{brookeSUSRetrospective2013}. % aimed at assessing system effectiveness, efficiency, and satisfaction \cite{brookeSUSRetrospective2013}. 
A SUS survey was conducted as part of our user study
detailed in \autoref{sec:userstudy}. %(see Section 6.2 of the paper for study details). 
Each of the 12 subjects completed the survey after completing their work with the system. The specific prompts used in the SUS survey and aggregated results are shown in \autoref{fig:sus}. 
We reverse the score of negative questions (so zero is the worst score for any question).
%
% The average overall score is 74.17. 
% The results suggest that our participants appreciate the system and thought they would use it. They also acknowledged its complexity. \sysname does take time to learn. We feel this is acceptable given the target of an experienced audience of researchers who can invest the time to learn the tools. If anything we were pleasantly surprised that study participants were not only able to achieve tasks, but also to identify unexpected ways to use the system even in the short time they had to work with it. 
%
The standard scoring method gives a score of 74.17 which is slightly above average. We feel this is acceptable given the prototype nature of the system and target of an experienced audience of researchers who can invest the time to learn the tools. The results suggest that our participants appreciate the system and thought they would use it. They also acknowledged its complexity; \sysname does take time to learn. However, even in the limited time, users discovered complex (and unexpected) workflows. 

\subsection{Nearest-Neighbor Overlaps}
% An important feature of \sysname is that it incorperates multiple embeddings and allows users to compare results between them. Motivating this usecase are experimental results concerning the differences in results produced by prominent embedding approches such as TF-IDF and Specter. 

In \autoref{sec:similarity} we note that the two provided embeddings are different. To quanitify this we computed the overlap for the top nearest neighbors over the entire corpus (here we report the Visualization corpus). \autoref{fig:embeddingoverlap}A shows for the top 10 nearest TF-IDF neighbors how many appear in the top $n$ list of the Specter list. The leftmost bar indicates that the top 2 Specter match (list of length 2) overlaps with only 0.47 of the top 10 TF-IDF documents on average. For the nearest neighbor (list of length one) the value is zero (indicating that the nearest Specter neighbor never appears in the top-10 list of TF-IDF). The rightmost bar of the chart indicates that the average overlap between the top 10 TF-IDF and top 100 Specter documents is 4.77, that is, on average, less than half of the top 10 recommendations appear in the top-100 list of the other metric. 
The lack of overlap may have positive benefits in terms of deduplication \cite{jannachExploringMultiListUser2021} and providing diversity \cite{puTrustinspiringExplanationInterfaces2007} but mixed impact on trust \cite{puTrustinspiringExplanationInterfaces2007}.

We tried to assess word matching explanations for nearest neighbors in cases where they are not causal. For TDIDF, similarity is caused by matching words, whereas for Specter matching contributes indirectly. \autoref{fig:embeddingoverlap}B shows results of an experiment with average word overlaps for neighbors across the visualization corpus. We see that TF-IDF and Specter neighbors share similar numbers of words, and that the number of shared words decreases for weaker matches (more distant neighbors). 

% In the context of \sysname embeddings are used to determine the most similar papers to a given selection. Through experimentation, it was found that TF-IDF and Specter produce very different results for this task. In \autoref{fig:embeddingoverlap}, we can see that the number of documents that both embedding approaches consider to be similar is on average very small, even when looking at the top 100 most similar documents for a given embedding. This is further supported by the fact that even the highest ranked documents in an embedding are frequently missing in the other. For about 30\% of the papers, the top ranked neighbor in Specter is not included in the top 100 neighbors in TF-IDF and about 37\% for the reverse. 

% Evidently embedding approaches are meaningfully different, however, attempted evaluation of which one is better proved to be inconclusive, as it is highly subjective and requires a high level of expertise in the domain of documents to do accurately. As such, we have included both embeddings in the application to allow for increased diversity of recommended papers and leverage the strengths of each.

\susstats
\embeddingoverlap

\end{document}